\documentstyle[aaspp4]{article}
\def\arcs{\char'175\ ~}

\def\hub{\ifmmode H_\circ\else H$_\circ$\fi}
\def\kms{~km~s$^{-1}$\ }
\journalid{337}{15 January 1989}
\articleid{11}{14}
\begin{document}

\title{The Butcher-Oemler Effect at Low Redshift: Spectroscopy of Five Nearby Clusters of Galaxies}

\author{Nelson Caldwell\altaffilmark{1, 2}}
\affil{F.L. Whipple Observatory, Smithsonian Institution, Box 97, Amado AZ 85645}
\affil{Electronic mail: caldwell@flwo99.sao.arizona.edu}

\author{James A. Rose\altaffilmark{2}}
\affil{Department of Physics and Astronomy, University of North Carolina, Chapel Hill, NC 27599}
\affil{Electronic mail: jim@wrath.physics.unc.edu}

% Notice that each of these authors has alternate affiliations, which
% are identified by the \altaffilmark after each name.  The actual alternate
% affiliation information is typeset in footnotes at the bottom of the
% first page, and the text itself is specified in \altaffiltext commands.
% There is a separate \altaffiltext for each alternate affiliation
% indicated above.

\altaffiltext{1}{Visiting Astronomer, Cerro Tololo Inter-American Observatory. 
CTIO is operated by AURA, Inc.\ under contract to the National Science
Foundation.} 
\altaffiltext{2}{Visiting Astronomer, Kitt Peak National Observatory.  KPNO is
operated by AURA, Inc. \ under contract to the National Science Foundation.}

\begin{abstract}
We present multi-fiber spectroscopy and broadband imaging of early-type galaxies in
five nearby rich clusters of galaxies.  
The main purpose was to look for ``abnormal'' spectrum 
galaxies (i.e., post-starburst galaxies which have
strong Balmer absorption lines and emission 
line galaxies) in nearby clusters that are
similar to those found by Caldwell et al. 
(1993) in the Coma cluster.  
Three of the clusters were purposefully selected to have pronounced double
structure in their spatial distribution of the galaxies.

Our primary conclusion is that
$\sim$15\% of the early-type galaxies in these nearby rich clusters
have signs of ongoing or recent star formation.
Furthermore, the starburst and post-starburst nature of these abnormal
spectra is reminiscent of that seen in distant clusters, although at a
reduced frequency and burst strength.  Thus activity
similar to that seen in distant clusters is still ongoing,
at a reduced level, in present-epoch rich clusters.
The frequency of such galaxies appears to
be enhanced significantly over that seen in field galaxies.
Most of the new post-starburst galaxies
are disk galaxies; three E galaxies in one cluster and one in another cluster
have current star formation.

We also find evidence in the spatial and kinematic structure of several of
the clusters that subclusters have recently passed through the
main clusters and are now emerging out the other side.  This inference is
based on a comparison of the spatial and kinematic cluster data with N-body
simulations of infalling clusters.  Specifically, we see evidence of velocity
gradients and/or a dispersed appearance of the observed subclusters,
both of which are produced by tidal distension and disruption of 
infalling subclusters.  If, as we suspect, the subclusters passed
through the main clusters more than a Gyr ago,  then the
post-starburst timescales of $\sim$1 Gyr imply 
that the star formation bursts are only triggered during (or after)
the subcluster passage through the center of the main cluster.  
We speculate that shocks
induced in the collisions of the cluster and subcluster ICM's may trigger the
galaxy starbursts.

\end{abstract}


\section{INTRODUCTION}

The influence of environment on the formation and evolution of galaxies
remains one of the central issues in the field of galaxy evolution. 
In that regard clusters of galaxies have long played a unique role in 
evaluating the importance of environment.
Perhaps the most dramatic example of the
impact cluster studies have had on our view of galaxy evolution is the discovery
by Butcher \& Oemler (1978) 
that a large fraction ($\sim$30\%) of galaxies
in the cores of distant rich clusters have abnormally blue colors. 
The ``Butcher-Oemler (B-O) effect'', originally described in Butcher \&
Oemler (1978) has been followed up by comprehensive broadband photometric 
studies (e.g., Butcher \& Oemler 1984; Couch \& Newell 1984; Dressler \& Gunn 1992) of a large number 
of clusters.  The principal result of those studies has been that many
galaxy clusters above a redshift of z$\sim$0.3 contain a large blue galaxy
population.

The crucial next step toward understanding the B-O effect
involved obtaining confirmation that the
blue galaxies are indeed cluster members, through
both narrow band photometry (e.g., Couch et al. 1983; Ellis et al. 1985; 
MacClaren et al. 1988; Rakos \& Schombert 1994; Belloni et al 1995) and spectroscopy 
(Dressler \& Gunn 1982, 1983, 1992; Lavery \& Henry 1986; Henry \& Lavery 1987;
Couch \& Sharples 1987; Soucail et al. 1988; Pickles \& van der Kruit 1991).
These studies have indeed confirmed membership for a majority of the 
blue galaxies.  Furthermore, the spectra have provided key information 
concerning the nature of the blue galaxies.  Specifically, many of them
exhibit strong emission line spectra, generally due to the presence
of active star formation, but in some cases due to an active galactic nucleus (AGN).
For the actively star-forming emission-line spectra, it is not 
in general clear whether the star-formation rates are typical of normal spirals or
unusually strong, since star formation rates in normal spirals are a strong
function of morphological type.  Thus in the absence of morphological 
information, it has generally been difficult to
establish whether a particular star-forming emission-line spectrum indicates
an unusual recent star formation rate.

On the other hand, a large number of the remaining  galaxy spectra, i.e.,
ones without detectable emission lines, definitely 
do have unusual star formation histories.  Specifically,
they exhibit strong Balmer absorption lines but no detectable emission, which
indicates that star formation has only recently ceased in those galaxies,
either because a major starburst has just ended or because ongoing star 
formation has been truncated recently.  These spectra, which were first
discovered by Dressler \& Gunn (1983), are commonly referred to 
as ``E+A'' or ``post-starburst'' (PSB) (e.g., Dressler 1987) and are 
relatively rare in the centers of 
nearby rich galaxy clusters, such as the Coma cluster.    Perhaps the most
surprising result to emerge from the spectroscopy is the realization that many 
of the photometrically red galaxies exhibit PSB spectra, thereby indicating 
that they have also recently experienced enhanced star formation (e.g., 
Couch \& Sharples 1987; Dressler \& Gunn 1992).  Thus the
fraction of evolving galaxies in distant clusters is actually higher than
originally inferred from the broadband photometry.

Ground-based images of several distant clusters by
Thompson (1988), Lavery \& Henry 1988, and 
Lavery et al. 1992 have demonstrated that a
considerable fraction of the B-O blue galaxies are in tidally interacting
systems, a result that has major implications for understanding the origin
of the blue galaxies.  Recently, HST has produced spectacular images 
of four distant clusters, which have greatly clarified some of the
central issues regarding the B-O effect (Dressler et al. 1994a,b; Couch et 
al. 1994; Wirth et al. 1994).  First, the images have demonstrated that most of
the blue galaxies are late-type spirals or irregulars, a galaxy population that
is virtually absent in present-epoch rich clusters.  In fact, Dressler et al.
(1994a) have proposed that the central issue concerning the B-O effect is the
disappearance of this large population of late-type galaxies from rich clusters
between z$\sim$0.3 and the present epoch.  Furthermore, a substantial
fraction of the blue galaxies are members of tidally
interacting or merging systems, thus confirming the ground-based studies cited
above.  Finally, Couch et al. (1994) find for their two clusters that the {\it
red} galaxies with PSB spectra are predominantly bulge-dominated systems, and
show little evidence for tidal interaction or merging.

Some central questions regarding the B-O effect are: (1) why is the star
formation that is prevalent in such a large fraction of galaxies at redshifts
of z$\sim$0.3-0.5 virtually absent at the present epoch?  (2) is the
fraction of starburst/post-starburst galaxies in nearby clusters truly zero; and
(3) what is the
mechanism (or mechanisms) which quenches that star formation, thereby leading to 
the post-starburst appearance of so many galaxies in the distant clusters?

These principal issues have been placed in a rather
different light by several studies of both early-type and late-type 
galaxies in nearby rich clusters.  Bothun \& Dressler (1986) studied several
star-forming spirals in the Coma cluster that appear to resemble blue
galaxies in distant clusters in several respects.  Gavazzi and
collaborators (Gavazzi 1989; Gavazzi et al. 1995), and Moss \& Whittle (1993) have shown
that while on the whole the spirals in nearby rich clusters are depleted in both
neutral and ionized gas, a surprisingly high fraction of the {\it early-type} 
spirals exhibit enhanced star formation rates in their central regions.  
Recently, Caldwell et al (1993; hereafter Paper I) have found that
many early-type galaxies in the SW region of the Coma cluster exhibit
spectroscopic evidence for recent bursts of star formation.
These ``abnormal-spectrum'' galaxies appear to coincide with a spatial
(Mellier et al. 1988) and kinematic (Colless \& Dunn 1996; Biviano et al. 1996)
substructure centered on the cD galaxy NGC 4839, as well as with a secondary 
peak in x-ray images of the Coma cluster (Watt et al. 1992; Briel et al. 1992;
White et al. 1993).
In a follow-up study
Caldwell et al. (1996) have shown that the recent star formation
found in the ``abnormal-spectrum'' galaxies,
while spatially extended over typically $\sim$2 kpc, is centrally concentrated,
as is the case for the Moss \& Whittle (1993) star-forming early-type spirals.
On the whole, the spectroscopic signatures found in the Coma cluster galaxies
are remarkably similar to those characteristic of many of the
distant cluster blue and red PSB galaxies.  

The existence of currently starbursting (SB) and PSB galaxies 
in nearby clusters clearly provides a different perspective on the B-O effect.
It also offers an unprecedented opportunity to unravel the
fundamental cause(s) of that effect, since the nearby clusters are more than ten
times nearer than the typical z$\sim$0.4 clusters.  In Coma one can
obtain substantially higher spatial reolution imaging and higher S/N ratio 
spectra, study galaxies farther down the luminosity
function (and thereby observe a much larger number of galaxies) than at z=0.4, 
and in so doing potentially obtain a variety of clues about the fundamental 
process(es) driving the B-O effect.
Thus it is important to determine whether
the high fraction of SB/PSB galaxies in the SW region of Coma is an anomalous
situation, or whether such activity is still widespread at the present epoch.
To that end, in this paper we present spectroscopy and photometry of early-type 
galaxies in five nearby rich clusters, including additional data for Coma
itself.  Three of the four new clusters (DC0103--47/0107--46, 
DC0326--53/0329--52,
and DC2048--52) have been selected for their pronounced
double structure, while the fourth (DC1842--63) is relatively regular
in appearance.  In \S 2 we describe the properties of the five clusters,
while in \S 3 we present the new spectroscopic and imaging data.  In \S 4 we
discuss the frequency, spatial distribution, and kinematics of the abnormal
spectrum galaxies in the five clusters, and a discussion of these results is
given in \S 5.  A summary and conclusions follows in \S 6.

\section{Observational Data}

\subsection{Spectroscopy}

\subsubsection{CTIO 4m Argus data}
Spectroscopic data were obtained for early-type galaxies in 
the four southern clusters mentioned above
in September of 1992 and August of 1993, using the Argus 24-arm positioner
and spectrograph, which is a facility instrument on the CTIO 4m telescope.
Input coordinates for the galaxies in these clusters were measured using either
the Grant measuring engine at CTIO, or the one at KPNO, using the 
Dressler (1980) coordinates or those of Malumuth et al. (1992) as first estimates.  The coordinates reported in the table below are accurate to about
1.5\arcs.  Observations were limited to the early-type galaxies as
classified by Dressler (1980), except as noted in section 3 below.
Because of the
density of targets in these clusters, several setups were required per field
observed.  Statistics on the numbers of galaxies observed relative to the 
number of candidates appear below.  

The fibers fed a bench mounted spectrograph, whose detector was a Reticon
1200X400 CCD.  A blue-blazed grating was used that gave a resolution
of 2 pixels or 3.6 \AA, and a wavelength coverage from 3500 to 5700 \AA.
Exposure times for setups in which most of the galaxies were bright were
one hour; those for fainter galaxy setups were from 3 to 5 hours long.
Raw frames were debiased, and flatfielded with a ``milk flat,'' which is
a frame taken with a plate of milky plastic between the output of the
fibers and the spectrograph camera.  Such a frame, which is an exposure of
the daytime sky, reveals the pixel-to-pixel sensitivity variations of the CCD
without having to worry about the fiber images being located in different 
places on the CCD chip in the flat than
in the nighttime data, as is often the case.  This frame had the spectral 
response modeled and divided out as well.

Separate images were combined in a way that removed the cosmic rays (see
Caldwell et al. 1996). The IRAF package ``doargus'' was employed for the
remaining reductions, which involved extraction, wavelength calibration,
fiber normalization, sky subtraction and flux calibration.  Standard
stars observed during the same nights provided the flux calibrations
and template stars for velocity measurements.

\subsubsection{KPNO 4m Hydra data}

The clear night of 1994 15 June was used to obtain KPNO 4m Hydra 
multi-fiber data of early-type galaxies in 
one more field in the Coma cluster, to add to the two 
that we reported on in Paper I.  Morphologies were obtained from 
Dressler (1980) or Caldwell et al. (1993).  The field was chosen to
be on the other side of the center of the cluster from the SW field
already observed, thus allowing us to check whether the high frequency
of abnormal spectra galaxies in the SW region was truly localized, or
a more general radial phenomenon.
The setup and data reductions
were the same as was reported in the earlier paper, with the exception that a
milk flat was employed with this data in a similar fashion as with the
CTIO data.  Fifty-four galaxy spectra were obtained; three of these are
background galaxies and two of them are late-type spirals, which we do not
include in later analyses of abnomral spectrum early-type galaxies.
Template stars were observed during the night for use in
determining velocities, but flux standards were not. Instead, the flux calibration
technique described in Paper I was employed.

\subsubsection{Miscellaneous spectral data}
The time between multi-fiber runs was sufficiently long to prompt us to
consider short cuts to finding the post-starburst galaxies in Coma. Dramatic
cases would of course have bluer than average colors, as we found for
the Coma PSB's reported in Paper I.  Thus a way to find further
PSB galaxies in Coma is to inspect the
color magnitude diagram from the Coma data of Godwin et al. (1983) for 
early-type morphology galaxies which have much bluer colors than the mean
relation would predict for the galaxy's luminosity.  There are 9 such
galaxies to be found in the GMP data.  Of those, five are already known to be
PSB or SB galaxies from Hydra spectra, and a sixth is 
reported on here (numbers 15, 43, 89, 94, 99, and 189 in Dressler's list).
The three others were observed with the MMT and the FLWO 1.5m telescope,
and one of these turned out to be a post-starburst galaxy (\#61;
\#85 and \#225 have normal spectra).  We include
it with the discussion of galaxies in the NE field in Coma, because
it could have been observed during the 4m Hydra observing run. The fact 
that it wasn't was simply due to the limited number of fibers available.

Coma \#89 was originally observed with the multi-fiber 
sample of Caldwell et al. (1993),
but the resultant spectrum was vignetted to the extent that the PSB nature of
the galaxy was questionable. An MMT spectrum of the galaxy does confirm that it
is a PSB, also in the SW field. The MMT spectra of \#61 and \#89 were taken
with the Blue Channel Spectrograph, and cover the region of 3500--5200 \AA \
with a resolution of 3 \AA.

\subsubsection{Radial velocities}

Heliocentric radial velocities have been derived for all of the multi-fiber
spectra obtained with the KPNO and CTIO 4-meter telescopes.  For all
galaxies with reasonably normal spectra (i.e., not dominated by emission
lines and/or hot stars), velocities were derived by using the {\it fxcor}
routine in the IRAF {\it rv} radial velocity package.  Two different KIII
spectra were used as templates in the cross-correlation determination.
The two templates gave very similar results.  As a check on the accuracy of
the Argus CTIO 4-m velocities, we have checked our heliocentric radial
velocities against those published by Malumuth et al. (1992; hereafter MKDFR) 
for DC0107--46
and DC1842--63.  For DC0107--46, there are 45 velocity measurements in common
between us and MKDFR.  After throwing out one galaxy with a
nearly 2000 \kms discrepancy (Malumuth \# 103, for which our spectrum
had a low S/N ratio), we found a mean velocity difference between us and
MKDFR of --3.2 \kms (in the sense that the MKDFR
velocities are slightly higher than ours) and an rms deviation from the fit
between the two sets of velocities of 108 \kms.  For DC1842--63 there are 14
velocities in common between MKDFR and us.  There is a mean
offset in velocity of only --2.4 \kms and an rms deviation from the fit to the
two sets of velocities of 64 \kms .  In Coma Colless \& Dunn (1996) have
compared 45 of their velocities against ours, and find a mean offset of
+22 \kms (with our velocities higher than theirs) and an rms scatter of
only 48 \kms .  Thus there is excellent agreement in both the mean and rms
scatter in our velocities with other determinations.

\subsection{Imaging}
We expect that  spectroscopic signatures of recent star formation in E/S0 
galaxies could
be accompanied by peculiarities in galaxy morphology or deviations in 
color from the normal E/S0 colors.  Consequently, we obtained CCD images
in B and R of three of these clusters, and have extracted magnitudes, colors,
and light profiles as well as morphological information from those.
Frames of DC2048--52, DC0103--47/0107--46, and DC0326--53/0329--52
were taken at CTIO in October and December of 1992 with the
0.9m telescope, during photometric weather.   Exposure times were
typically 10 minutes in Kron-Cousins R and 20 minutes in B.  
The majority of the 
galaxies in those clusters for which we have 
spectra were imaged with the CCD, but a few were not.
Standards were reduced by S. Tourtellotte of Yale. 

The galaxy photometry was extracted as follows.  An initial estimate was
made of the sky around the galaxy being measured, and subtracted from the
frame.  A program which fits ellipses to the isophotes was run, out to a
radius about twice that of the expected limiting radius of the galaxy.
A correction to the sky level was then made, based on the residual intensity
measured at large radii.  Having found the R band isophotes, the B band data
was then extracted using the R band ellipses, adjusted for a change
in the centering of the galaxy.  
Magnitudes and colors as a function of radius were calculated
using both sets of isophotes now.  We elected to measure the R 
magnitude enclosed
within an R surface brightness of 26, and B--R colors within r=r(R$_{26}$)/2.
Data for abnormal spectrum galaxies were checked for color gradients, as
this might be expected in post-starburst galaxies, but in general those
galaxies were so faint that the current data is of little use in measuring
such.

Errors in magnitudes and colors were determined from
a combination of the errors from photon noise in the object and sky,
readout noise, mean error of the sky level, and the transformation to
the standard system.  Several galaxies appeared on more than one frame, 
and were used to test the repeatability of the reduction process.  The
instrumental magnitudes and colors from the two frames for these galaxies
agree within 0.05 and 0.02 mags respectively.

To compare the clusters, we have made reddening and K corrections and
converted the magnitudes to an absolute scale.  For DC2048--52 
(z=0.045), we used
k-corrections K(R)=0.04 and K(B--R)=0.20, and extinction corrections
A(R)=0.03 and E(B--R)=0.02, as derived from the K-corrections of Coleman, 
Wu, \& Weedman (1980), and the
reddening maps of Burstein \& Heiles (1982). Likewise, for DC0326/0329
(z=0.058), K(R)=0.05, K(B--R)=0.26, and E(B--R)=0.0; for DC0103/0107
(z=0.025), K(R)=0.02, K(B--R)=0.11, and E(B--R)=0.0.  \hub \ was taken to
be 80 \kms Mpc$^{-1}$ and q$_\circ$=0.1. 

Finally, images of abnormal spectrum galaxies were inspected for peculiarities, and the 
light profiles of all were classified by eye as to whether they were 
exponential or more R$^{1/4}$ in nature, indicating whether the galaxies are
disklike or spheroidal. 

\section{Selection of Clusters of Galaxies}

In this Section we provide a brief description of the basic properties of each
of the galaxy clusters studied and the rationale for selecting them. Table 1
contains summary information about the morphological mix of the clusters and
the observations that we report.  All of these clusters are spiral poor.
DC0326--53/0329--52 contains an exceptionally 
high number of galaxies classified as ellipticals, although Oemler (1992)
does report some other clusters with such a high proportion.  
We note that DC0326--53/0329--52 is also
the most distant cluster in our sample and hence the morphologies might
be systematically in error.  We return to this subject later in \S 5.

1) {\bf Coma.}\\
In the study of Paper I it was shown that there is a high
proportion of abnormal spectrum galaxies in the SW region of the Coma cluster.
Due to limited clear weather, we were unable to observe galaxies in other 
sectors of the Coma cluster, other than in the central region.  Thus an obvious
question that needs to be addressed is whether the high number of abnormal
spectrum galaxies is strictly a phenomenon of the SW region of Coma or is
widespread throughout the cluster.  To that end we have observed a number of
galaxies in a region on the NE side of the cluster, i.e., opposite to the
SW region where the abnormal spectrum galaxies appear to be concentrated.
To illustrate the current state of our spectroscopy in Coma, in Fig. 1 are
plotted all of the galaxies observed spectroscopically in Paper I (central
and SW regions) and here (NE region) along with the remaining galaxies in the
Godwin et al. (1983) catalog down to a limiting magnitude of B=17.5.

While it was originally proposed that nearby rich clusters are dynamically 
relaxed systems, more recent studies have demonstrated that
present-epoch rich clusters frequently contain one or more spatial/kinematic
substructures, which indicates that they are less dynamically developed than
had been assumed (e.g., Baier 1983; Dressler \& Shectman 1988b; 
Beers et al. 1991; Bird 1994).
In particular, there is strong evidence for the presence of several 
spatial/kinematic substructures in the Coma cluster, with the most prominent one
appearing in the SW region approximately coincident with the abnormal spectrum
galaxies (E.g., Mellier 1988; Colless \& Dunn 1995; Biviano et al. 1996).
Thus it is natural to suspect a connection between the high
incidence of abnormal spectrum galaxies in the SW region and the spatial/kinematic
substructure there, particularly if (as will indeed be demonstrated in \S 5) the
incidence of abnormal spectrum galaxies is lower in other areas of the cluster.
To further investigate the possible connection between star-formation activity
in cluster galaxies and the presence of substructures, we have studied
four other clusters with various degrees of substructure.

2) {\bf DC2048--52.}\\
The southern cluster DC2048--52 (ACO 3716) was studied by Dressler (1980), who
gave magnitudes and morphologies for 247 galaxies in the cluster.  From
Dressler's study it is clear that the cluster has a very pronounced northern
substructure, to the extent that it can almost be seen as a double cluster.
However, as can be seen in 
in Fig. 2, where the galaxies observed spectroscopically by us are plotted
as filled circles, and remaining galaxies in Dressler's catalog are plotted as
filled circle, the central cluster is more populous and centrally concentrated
than the northern substructure.  Dressler \& Shectman (1988a) published
velocities for many galaxies in DC2048--52.  They (Dressler \&
Shectman 1988b) demonstrated that a large
velocity offset exists between the main cluster and the northern substructure.
In addition, the velocity dispersion of the main cluster is
substantially higher than that of the northern substructure.

3) {\bf DC0326--53/DC0329--52.}\\
The double cluster DC0326--53 and DC0329--52 (ACO 3125 and 3128 respectively)
represents perhaps an even more 
extreme example of substructure than DC2048--52.  It is in fact catalogued
as two clusters by Dressler (1980).  The distribution of galaxies in the
two structures is plotted in Fig. 3.  There the DC0329--52 component is
seen to be more centrally concentrated and richer than the more extensive
and diffuse DC0326--53 component that fans off to the SW of DC0329.  No
extensive radial velocity data for the cluster existed prior to this study.

4) {\bf DC0103--47/DC0107--46.}\\
The double cluster DC0103--47 and DC0107--46 (ACO 2870 and 2877 respectively)
provides yet another excellent example
of pronounced substructure.  As in the case of DC0326/0329, one component
of the ``double cluster'', DC0107--46, is substantially richer, more centrally
concentrated, and more symmetrical than the other, DC0103--47.  DC0103--47 
itself appears to consist of subclumps (to be discussed further in \S 5)
and generally fans off to the SW of DC0107--46.  This can be seen in Fig. 4,
where the distribution of galaxies studied by Dressler (1980) and MKDFR is plotted.
They found evidence for several substructures, including a low velocity 
structure in the general area of DC0103 and a low velocity component near
the center of DC0107.  Perhaps their most interesting result is the 
identification of a background sheet of 18 galaxies, at a redshift of
$\sim$9000 \kms, that extends over the entire area of DC0107 and
into DC0103 (see their Figs. 4 and 7).  

5) {\bf DC1842--63.}\\
In contrast to the complex structure of DC0103/0107, the cluster DC1842--63
supplies a counterexample of a cluster with no clear evidence for
substructure.  Bell \& Whitmore (1989), and later
MKDFR also extensively studied DC1842--63,
the latter group finding on the basis of 154 radial velocities  
no indication for substructure.

\section{Definition of Abnormal Spectrum Galaxies}

\subsection {Spectral Analysis}
In Paper I many of the spectra of early-type galaxies in the SW region of
the Coma cluster were classified as ``abnormal''.  An ``abnormal'' spectrum 
for an early-type galaxy was considered to be any spectrum showing signs of 
recent star formation and/or an active galactic nucleus (AGN).  Specifically, 
enhanced Balmer absorption lines 
provided one quantitative measure of recent star formation.  Alternatively,
non-AGN emission lines were used as evidence for {\it ongoing} star formation.
Emission spectra with line intensity ratios different from those in standard
HII regions (e.g., high excitation coupled with high 
[NII]$\lambda$6584/H$\alpha$) were considered to be AGN spectra.  However,
while it was demonstrated in Paper I that the distribution
of galaxies with high Balmer absorption line strengths in the 
Coma SW region is significantly
different from that of galaxies in the central region, in individual cases the decision
as to whether a galaxy spectrum could be considered abnormal was based on a
subjective evaluation of the entire spectrum.  Here we define a quantitative
criterion for an ``abnormal'' spectrum galaxy that provides an objective
and reproducible assessment of which galaxies should be considered as
abnormal and that also agrees well with our subjective evaluations of the
spectra.

We first return to the qualitative definition of an abnormal spectrum galaxy.
For an early-type (E or S0) galaxy, {\it any} sign of recent star formation
is clearly abnormal.  However, if we consider spiral galaxies,
then the situation becomes more
complicated, since spirals in general have a certain amount of ongoing
star formation.  In that case we must generalize the definition of abnormal to
be any galaxy with an {\it unusually} high current or recent star formation rate
and/or unusually strong AGN emission.  This immediately begs the question of
what constitutes an ``unusual'' current or recent star formation history
in galaxies of various morphological types.  As a benchmark for what the
spectra of normal galaxies are like as a function of morphological type we
use the library of global spectra of galaxies published in Kennicutt (1992a,b).
As was 
mentioned in the Introduction, the PSB spectra (i.e., enhanced Balmer
lines with no detectable emission) are not seen at all in ordinary spirals;
any normal spiral with signs of star formation generally has detectable 
H$\alpha$ \ or [OII] $\lambda$3727 emission to indicate that the star formation is ongoing.
Thus we can generalize that {\it any} galaxy, regardless of morphological type,
with an PSB spectrum can be considered to have an unusual recent star 
formation history.  A spectrum with Balmer absorption but no emission lines
implies either a recently terminated star formation burst or recent
truncation of a steady past star formation history, which in either case is
unusual for a galaxy.

The more confusing situation arises in those cases where the enhanced Balmer
lines are accompanied by detectable emission, hence implying ongoing star
formation (and/or an AGN).  Based on the data for the
equivalent width of [OII]$\lambda$3727 emission for the normal spirals in the 
Kennicutt (1992a,b) sample, we propose that an unusual star formation rate for
an Sa galaxy requires an equivalent width in [OII]$\lambda$3727 in excess of
5 \AA, while for an Sb and an Sc the equivalent width must exceed 20 \AA \ and
40 \AA \ respectively.  Thus, if we allow for an error of one Hubble type in
the morphological classifications, then we must allow for the possibility
that an E or S0 galaxy is actually a mis-classified Sa.  In that case we
can consider as definitely abnormal any spectrum with enhanced Balmer
absorption and with [OII]$\lambda$3727 emission that is either undetectable
(PSB scenario) or in excess of 5 \AA \ (excessive for a normal Sa).  

In Paper I galaxies with enhanced Balmer absorption lines were
ultimately picked out from a subjective examination of the spectra.  
Nevertheless, quantitative measures of the Balmer line strengths were utilized
in the paper.  Specifically, an equivalent width of H$\delta$ was defined,
along with the CN-H8 slope index that in effect measured the strength of H8 
relative to the CN $\lambda$3883 molecular band.  While these measures were
used to demonstrate that the distribution of Balmer line strengths is 
different for the early-type galaxies in Coma SW compared to those in the
central region of the cluster, in individual cases the subjective examination
of the whole spectrum was given greatest weight.  The primary advantages of the
subjective approach were (1) all Balmer lines could be used and weighted
according to their quality and (2) the S/N ratio of the spectrum could be
evaluated in arriving at a conclusion as to whether or not a spectrum should
be considered abnormal.  In this paper we have defined a quantitative criterion
for a post-starburst spectrum that is designed to mitigate these
difficulties.  

The H$\delta $ equivalent width and the CN/H8 
slope index already mentioned are used, along with a simple equivalent width
of the H8 line and the same for the K line. The precise windows for
the index measurements are listed in table 2 (the definition of CN/H8 is
contained in Caldwell et al., 1993).
The H8 equivalent width will have meaning
only when the line dominates over the normally present CN line, and as such
serves to strengthen the confidence in the CN/H8 line.  The equivalent width 
of the K line diminishes
in the presence of a young population.  As we define it, the line has a strength of
between 10-13 \AA \ in normal galaxies, but drops to 6-9 \AA \ in post-starburst galaxies.
In principle, color could also be used, but we do not have the same confidence in
the derived continua slopes for our spectra as we have in the line strengths.

The lines are combined into a significance value D as follows.  Using the flux calibrated,
deredshifted data, we measure the 4 indices for all spectra, including the spectra of
the central region of the Coma cluster in Paper I.  The indices for the
normal galaxies in the central Coma region are used to derive both the mean and 
standard deviation for each index. Now all those indices are normalized by subtracting the
appropriate means and dividing by the appropriate standard deviations.  These scaled values
are then combined in a way that maximizes the sensitivity to the presence of young
stars (H$\delta $ and H8 are added, while CN/H8 and K are subtracted).  

\noindent
(1) ${\rm d}_i= {({\rm H}\delta_i-\mu_{{\rm H}\delta}) \over \sigma_{{\rm H}\delta}}
+ {({\rm H8}_i-\mu_{\rm H8}) \over \sigma_{\rm H8}}
- {({\rm K}_i-\mu_{\rm K}) \over \sigma_{\rm K}}
- {({\rm CN/H8}_i-\mu_{\rm CN/H8}) \over \sigma_{\rm CN/H8}}$

The variance S$^2$ for these normal spectrum d$_i$ values is
then determined using the signal/noise ratio of the included spectra as 
weights (the mean of the d values is already zero).

\noindent
(2) ${\rm S}^2 = {\sum({\rm w}_i*{\rm d}_i^2) \over \sum{\rm w}_i} * ({N \over N-1})$,

where w$_i$ are the weights, and N(=59) is the number of Coma normal galaxies 
(Bevington \& Robinson 1992).
The variance S$^2$ for the normal galaxies is then used to determine
the significance D of the deviations for an individual spectrum in all of
our clusters from the mean indices for 
the Coma normal spectra:

\noindent
(3) ${\rm D}_i^2 = {\rm d}_i^2/{\rm S}^2*({\rm w}_i/{\rm W})$

where W is the mean weight of the Coma normal spectrum galaxies.  Thus the way
that D is defined, a spectrum with a D value of 2 deviates by 2 $\sigma$ from 
the typical ``normal'' Coma cluster early-type galaxy spectrum in the sense of 
having stronger Balmer lines and a weaker Ca II K line.
For instance, the post-starburst galaxies
\#43 and \#99 in Coma (Paper I) have D values of 7.5 and 5.4, respectively.
In weighting the D values by the S/N ratio of the spectra, we have opted
for maintaining a uniform quality of significance in the detection of
abnormal spectra at perhaps the expense of excluding some abnormal spectrum
galaxies which have only low signal-to-noise spectra.  

The D values for the normal Coma cluster spectra are 
correlated with their B magnitudes from GMP. This correlation
is simply another manifestation of the well-known color-absolute magnitude
relation for early-type galaxies (e.g., Sandage \& Visvanathan 1978;
Larson et al. 1980) and of its spectroscopic
(line-strength versus absolute magnitude) counterpart (e.g., Faber 1972, 1977).
Specifically, we find a slope of 0.275 in the sense that D becomes
larger at fainter magnitudes. Similar relations exist for the other clusters
(where the B magnitude data comes from our CCD photometry),
but because the data is less comprehensive than that for
Coma, the Coma fit was considered more reliable and was used for all
clusters after modifying the zero point to account for the different
distance moduli. To more
reliably identify the abnormal spectrum galaxies, we have 
subtracted off this standard D versus B relation, thereby producing
final ``corrected'' D values.  

In practice, there is a very good correlation between our subjective
determinations of ``abnormal'' spectrum galaxies (based on a visual
examination of the spectra) and those spectra with D values greater than 2.
Specifically, for galaxies in the Coma cluster brighter than B = 17.5,
agreement between the subjective and D value methods is found in more than
80\% of the cases where either the subjective or D value methods indicates an 
abnormal spectrum.  In general, in cases where the S/N ratio exceeds
$\sim$15:1 per 1 \AA \ pixel, there is good agreement between the two methods.
Thus for those cases where the visual inspection and the D value concur
in categorizing the spectrum as ``abnormal'', that designation is made.  In
cases where the two methods are not in accord, we turn to the photometric
color data for the galaxy, if available, for additional information.  

\subsection{Photometrically Blue Galaxies}

As was mentioned above, it is well known that early-type galaxies in rich 
clusters manifest a color-absolute magnitude relation similar to that found for 
field E and S0 galaxies.  The deviation of many galaxies in distant 
clusters from this standard c-m relation forms the original basis for the 
B-O effect.  
In Fig. 5 is plotted the c-m relation for all of the early-type galaxies that
are members of the 
three clusters for which we have CCD photometry.  A reasonably well defined
c-m relation is present, along with a number of galaxies scattered blueward
of this relation, indicating unusal star formation histories in those
early-type galaxies.  The solid line represents a linear fit to the galaxies
with normal colors.  We then measured the deviation of each galaxy with
respect to the mean c-m relation, in the sense that a blue galaxy is defined
to have a negative $\Delta$(B--R) color excess.  In Fig. 6 are plotted the
measured D values of the spectra versus the $\Delta$(B--R) color excess for
galaxies in DC2048--52.
It can be seen that for D $>$ 2, there is an excellent correlation between high
D value and large negative color excess.  Thus the global color data supplies
an additional characterization concerning the normal or abnormal nature of
the galaxy.  In those cases where a discrepancy exists between the visual
inspection of a spectrum and its D value, the $\Delta$(B--R) was used as
a tie-breaker in deciding whether the spectrum is abnormal or not.  In
particular, a cutoff at $\Delta$(B--R) less than --0.04 was chosen as the
boundary between normal and abnormal.

\subsection{The Abnormal Spectrum Galaxies}
The spectra identified as abnormal are shown in Fig. 7-11, with accompanying
CCD images (if they exist) in Fig. 12.  For the double cluster DC0326--53/0329--52,
we have used the Dressler (1980) numbering scheme, with suffix ``a'' if the galaxy
was in DC0329--52, and ``b'' if from DC0326--53. Likewise for DC0103--47/0107--46 the
suffix ``a'' refers to Dressler's list for DC0103--47 and ``b'' to DC0107--46. 
Galaxies in DC0103--47/0107--46 or in DC1842--63 with a suffix ``m'' were taken
from tables in Malumuth et al. (1992).

DC2048--52
appears to have a number of genuine PSB spectra, while DC0326/0329 has three
E galaxies with current star formation, but few PSB galaxies.  Those E galaxies
were classified as such by Dressler (1980), which are confirmed by our CCD
images.  One of those (DC0326--53\#80) is somewhat asymmetric, however, 
possibly indicating a recent encounter. 
We have indentified a Sy 2 galaxy in 
that cluster as well (\#54a).  DC0103/0107 has a number of PSB
galaxies, and one S0 with current star formation.  There is one weak PSB
galaxy in DC1842--63, and one E with a spectrum that essentially looks like that of
a spiral.

\#61 in Coma
is perhaps the strongest PSB in Coma, in the sense of the strength of its
Balmer absorption, though it is true that some residual star formation 
still exists.  The galaxy is an S0, which has a normal appearing inner part, 
but a faint asymmetric envelope.

In fact, most of the definite PSB cases in these clusters are S0 galaxies,
as evidenced by their morphological classes and their light profiles.
(Interestingly, HST studies of SB \& PSB galaxies in z $\sim$ 0.5 clusters
show that most such galaxies are also disk galaxies, Dressler 1996.)
The only unusual E galaxies are one PSB in DC0107--46 (\#30),  and
the three galaxies in DC0326/0329 and one in DC1842--63 with current star
formation. The DC0326/0329 galaxies  are also all quite blue, and 
at M$_{\rm R} \sim  -20 $ among
the least luminous galaxies in that cluster that we have studied.  It is 
possible that these are related to blue compact dwarf galaxies, though their
emission line strength is fairly weak for that class.  Their true
nature will have to be discerned with higher resolution images than we have
here.

Several of these galaxies have close companions, but there are no obvious
effects of interactions in these images.

\subsection{Summary of Data}

The results of the spectroscopic and imaging data for the four southern clusters are 
summarized in Tables 3-7.  Column 1 contains the designation given in 
Dressler (1980). Malumuth et al. (1992)
numbers were used for galaxies not listed in
Dressler (those with suffixes ``m''). Morphological types are from Dressler,
except for the Malumuth et al. galaxies for which types were determined by
the authors.  The type of profile (``e'' for R$^{1/4}$ profiles, ``d'' for
exponential profiles) is listed in column 5, and were determined from the
light profiles described above.  Other parameters gleaned from the CCD
images listed in the succeeding columns are: the ellipticity, the radius
at which the R surface brightness falls to 26 mags arcs$^{-1}$ (r$_{26}$, 
geometric mean of
major and minor axes), the R magnitude within r$_{26}$, the B--R color
within  r$_{26}$/4, and the deviation from the mean M$_R$--(B--R)$_{0,K}$
relation defined by the three clusters described above.  These numbers are
absent for the DC1842--63 for which we had no photometry; the Dressler
magnitudes are substituted.  The next column lists the 
heliocentric velocities, followed by the spectral signal-to-noise ratio,
determined at 5100\AA .  The equivalent widths of the K line, the H$\delta $
line and the H8 line, as well as the CN/H8 index are listed next, followed
by the D value.  The equivalent width of any [OII]$\lambda$3727 emission follows. The notes
describe the individual abnormal spectra (wPSB means weak post-starbust galaxy).

The table for the Coma cluster data (table 7), obtained NE of the center, is similar to the
previous. The GMP number, column 1, is from the survey of Godwin et al. (1983).
That reference also supplied the coordinates as well as the magnitudes and colors,
which are on the Johnson system rather than the Kron-Cousins system, hence the
B--R colors are 0.2-0.3 mag redder than the corrected colors 
in the other clusters for which we have CCD data.  The $\Delta $(B--R) colors
were obtained by finding the (B--R)--B relation for the normal spectrum galaxies in the
central field of the cluster, and finding the deviation from that relation for
the galaxies in question.

\section{Results}

\subsection{Frequency and Spatial Distribution of Abnormal-Spectrum Galaxies}

\subsubsection{Distribution}
While in Paper I it was found that an unexpectedly high fraction of galaxies 
in the SW region of the Coma cluster have recently undergone starbursts,
two key issues remained unresolved.  First, is the high fraction
of SB/PSB galaxies in the SW of Coma confined to that region of the cluster, or
is it characteristic of a general radial increase in the abnormal spectrum
galaxies?  Although the close coincidence between the location of the abnormal
spectrum galaxies with a spatial (Mellier et al. 1988) and kinematic (Colless \&
Dunn 1996; Biviano et al. 1996) substructure (and with the secondary
x-ray peak as well) offers cirumstantial evidence to support the hypothesis that
the SW region of Coma is unique in the cluster, spectroscopy of galaxies at a
similar radial distance from the cluster center but at other position angles
is clearly required to resolve this question.  Second, is Coma unique among 
nearby rich clusters in terms of its high SB/PSB galaxy content?
We now make use of our new spectra to show that the SW region in Coma is indeed
special to that cluster, and that several other rich clusters also contain 
abnormal spectrum galaxies.

In Fig 13 are plotted the locations of all Coma galaxies {\it with B$\le$17.5}
for which we have obtained multi-fiber spectroscopy.  The data in the central 
and SW regions
of Coma have been previously reported in Paper I, while the data for the
NE sector is new.  Galaxies with normal spectra are plotted as open triangles,
while those with abnormal spectra are plotted as filled triangles.
We have plotted 
only galaxies brighter than B of 17.5 to restrict the sample to spectra of
sufficiently high S/N ratio that identification of abnormal spectrum galaxies
is relatively unambiguous.  An inspection of Fig. 13 reveals that the
fraction of abnormal spectrum galaxies is much higher in the SW region than
at a comparable radial distance in the NE.  Thus the high abnormal
spectrum fraction in the SW region does indeed
appear to represent a unique situation in the Coma cluster.  On the other
hand, there are a few SB/PSB galaxies scattered around the central and NE
regions of Coma, indicating that the SB/PSB galaxies are not {\it exclusively}
located in the SW region.  These galaxies can perhaps be explained as being 
unrelated to those in the SW region (if they are individual galaxies falling into
the cluster for the first time, for instance), or as galaxies that once were
part of the group now seen in the SW but which have been stripped from that
group during its passage through the main body of the cluster (see \S 6).

To assess whether other nearby rich clusters contain similar abnormal spectrum
galaxies to Coma, we plot in Fig. 14 the locations of all galaxies in
DC2048--52 studied spectroscopically by us, with different plot symbols
reflecting whether the spectrum is normal or abnormal.  Overall, the fraction
of abnormal spectrum galaxies in DC2048--52 is seen to be roughly comparable to 
that in Coma.  However, the abnormal spectrum galaxies are much more evenly
distributed over the DC2048--52 main cluster and large northern substructure
than is the case for Coma, where most of the abnormal galaxies at the present time
are concentrated in the SW region.  Furthermore, while in Coma nearly all of the
abnormal spectrum galaxies are of the PSB variety, in DC2048--52 only about half
(or perhaps less) of the galaxies are PSB, the others showing emission lines
characteristic of ongoing star formation and/or AGN behavior.

A similar result generally applies to the other southern clusters.  
In Fig. 15 are plotted the locations of all galaxies
in DC0326--53/0329--52, again with symbols distinguishing the normal from abnormal 
spectrum galaxies.  The same plot is given for DC0103-47/0103-46 in Fig. 16,
and for DC1842--63 in Fig. 17.  For both DC0326--53/0329--52 and DC0103-47/0107--46
there is a substantial
fraction of abnormal spectrum galaxies, and the latter galaxies
are spread widely throughout the main cluster and associated substructure,
rather than being mostly confined to a specific region in the cluster.
In the case of DC1842--63 the number of galaxies observed is too small to draw
substantive conclusions, but the presence of 2 abnormal spectrum galaxies 
among the 17 observed is certainly consistent with the fraction of abnormal 
spectrum galaxies found in the other clusters.  Hence we conclude that the
phenomenon of abnormal spectrum galaxies is not unique to the Coma cluster
and instead is fairly ubiquitous among nearby rich clusters.  

\subsubsection{Frequency}
In Table 8 we
summarize the data on abnormal spectrum galaxies in each of the clusters and
their substructures.  Overall the fraction of abnormal spectrum galaxies is
typically $\sim$15\%, with approximately half of them being of the PSB variety.
However, the abnormal fraction varies from as high as $\sim$40\% in the SW region
of Coma to as low as $\sim$5\% in the central and NE regions of Coma.  In 
addition, the fraction of abnormal spectrum galaxies that are PSB varies 
considerably from one cluster to another.  While
in Coma nearly all of the abnormal spectrum galaxies are PSB, in 
DC0326--53/0329--52 only two of the 12 definite abnormal spectrum galaxies can
be classified as PSB (based on low or non-existent emission and enhanced Balmer
absorption lines).  Of the remaining 10 abnormals, only one has a high-excitation
spectrum characteristic of Seyfert 2 activity (54a), while the others have
emission spectra that are most likely due to current star formation, although
a LINER spectrum cannot be ruled out.  To further investigate this, Jon Morse
kindly acquired spectra of both \#30 in DC0329--52 and \#80 in DC0326--53.  The
spectra had enough coverage in the red so that H$\alpha$ was at the edge of the
Loral 1KX3K CCD detector.  From these spectra it is evident that the ratio of H$\alpha$ to 
[NII] $\lambda$6548 is high, and that [OI] $\lambda$6300 is absent.  Both of
these diagnostics indicate that the emission spectrum is characteristic of an
HII region, and thus of ongoing star formation.  In short, Coma appears to
represent one extreme, in which most of the abnormal spectrum galaxies
are PSB, while DC0326--53/0329--52 appears to be at the other extreme, where
pure PSB galaxies are in the minority.  DC2048--52 and DC0103--47/0107--46 lie
between these extremes.  The fact that so many {\it currently} star forming galaxies 
are present in DC0326--53/0329--52 makes it a particularly interesting cluster.

How do these frequencies compare with those seen in the more distant clusters, where
the Butcher-Oemler effect is in full splendor?  Since we limited ourselves to
observing early-type morphology galaxies 
(whereas the distant cluster observations had no
such restriction), we were relatively 
biased against spirals undergoing starbursts, and thus
the only pertinent comparison to be made is with the PSB
galaxies. This is itself difficult because of the difference in the
quality of spectra of the nearby and distant samples (we are able to find weaker
PSB cases due to our higher S/N).  
With that in mind, Dressler \& Gunn (1992) report that
17\% of members in seven z $\sim $ 0.5 clusters are PSB's
(and 21\% of non-emission line galaxies are PSB's).  
In their study of 3 clusters at z$\sim $0.3, Couch \& Sharples (1987) 
identified somewhat weaker cases of PSB (referred to as ``H$\delta $
strong'' as well as strong PSB galaxies.  Nearly
40\% of the member galaxies fall in those categories (and 50\%
of the non-emission line galaxies).
These values are clearly higher than the
nearby cluster PSB frequencies of 3-12\%, (which are the ratio of
PSB's to all non-spirals). Thus we may conclude 
that the incidence of PSB galaxies in clusters indeed declines 
with redshift (as expected within the context of the B-O effect), but it
is by no means zero for nearby clusters.

\subsubsection{Frequency of PSB galaxies in different environments}
Zabludoff et al. (1996) have identified  21 PSB galaxies out of 
approximately $10^4$ good spectra from 
galaxies with $0.05 < z < 0.13$ in the LCRS survey.  Of these, about 75\% 
are not in dense aggregates of galaxies,
leading them to conclude that the physical process that creates PSB galaxies is
not related to cluster environments, though of course there remains the
possibility that a different process may operate in the field than in clusters.  The 
key to further understanding is evidence concerning the {\it fraction} of PSB galaxies
in clusters and in the field.  Zabludoff et al. find that about 9600 galaxies are
not near clusters with internal velocity dispersions greater than 400 \kms.   There
are 16 PSB galaxies with similar characteristics, giving a frequency of
0.0017.  Oemler (1994) reports that only 30\% of field galaxies are early-type,
so assuming that PSB galaxies are early-types, the 
field frequency of PSB galaxies among early types is probably 
0.0017/0.30=0.0056 (16 out of 2880, for statistical purposes).  

It is possible that the LCRS undersampled cluster populations because of the 
technical difficulty of observing galaxies very close together on the sky
with fibers.  We therefore must look to other data sets such as ours
to find what the frequency of PSB galaxies is for nearby clusters.
To compare the Zabludoff et al.
number with the frequency we are finding for clusters, we should be sure that we
are using the same criteria for labeling a spectrum a PSB spectrum.  To that end,
we obtained a spectrum of one of the Zabludoff et al. galaxies (\#13 in their list)
with the MMT, and have measured the D value as well as the individual line
strengths for it following the prescriptions listed above.  Its D value is 4.5,
which is similar to the Coma galaxy \#99. Visual comparison of the two spectra is
gratifying in that the spectra do appear similar.  Zabludoff et al. defined
PSB galaxies to have an average strength of H$\delta $, H$\gamma $, and H$\beta $ 
greater than $<{\rm H}>$ = 5.5 and 
an equivalent width of [O II]$\lambda$3727 less than 2.5 \AA .  Their galaxy \#13 has an
$<{\rm H}>$ of 6.4, thus allowing us a crude conversion of the cutoff value from their
data set to ours. Thus we will assume that galaxies with D values greater than 
3.9 and low [O II] emission are of similar PSB strength to the Zabludoff et al.
galaxies.

There are 6 such strong PSB galaxies in the Coma cluster, 4 in DC2048--52, and
none in the other clusters.  The frequency of PSB's in Coma is thus
0.034 (6 out of 172), and 0.04 in DC2048--52 (4 out of 91).  
Over all five clusters it is 0.025 (10 out of 407).  Using statistics for
comparing two proportions (Miller et al. 1990), we find that the frequencies
for Coma, DC2048--52 and all five clusters together differ from the Zabludoff
field frequency by 4.41, 4.42, and 4.05 standard deviations respectively; i.e., the
field and cluster ratios would occur in a random selection process from the same parent
distribution only 6 times out of $10^5$.  

The cutoff applied by Zabludoff et al. is of course somewhat arbitrary. A more
lenient value would easily triple the number of PSB's in the LCRS sample.
Such a change could decrease the significance of the difference
between our clusters and Zabludoff et al.'s field, but because more cluster
galaxies are now included the change is not great.  For Coma, the significance
of the differences in the ratios changes to
4.37 sigma, and for all our clusters it changes to 2.78 sigma.

There are several possibilities for this discrepancy in frequencies. First, it
may still be an artifact of the data. We may have sampled the luminosity function
in these clusters to a much greater depth than was done in the LCRS field sample,
and if PSB galaxies are more common in less luminous galaxies, we may have
artificially increased the discrepancy.  This does not appear to be the case,
however. Neither we nor Zabludoff et al. report a luminosity dependence of
the frequency of PSB galaxies.  Second, the cluster environment may still
be the preferred though not exclusive place for forming PSB galaxies,
especially when one considers that PSB galaxies occur 40-80 times more
frequently in the z $\sim $ 0.5 clusters than in the nearby field. 

This would
seem to imply that more than one physical method is present for forming
PSB galaxies.

\subsection{Kinematics}

\subsubsection{Kinematic analysis}

Recently, kinematic studies of clusters of galaxies have advanced far enough 
in terms of the quantity and quality of multi-fiber and multi-aperture
spectroscopy that a more comprehensive look at cluster kinematics is 
now possible.
These studies have generally produced a picture in which nearby rich clusters
are still far from fully virialized.  Instead, evidence for kinematic (and
spatial) substructures has accumulated in several clusters.  In particular,
recent analyses by Colless \& Dunn (1996; hereafter CD) and Biviano et 
al. (1996; hereafter B96)
of hundreds of galaxy radial velocities in Coma have both
revealed what appears to be a velocity gradient in the main body of Coma, a
gradient that parallels the NE-SW direction of elongation of
the cluster.  Since the number of new velocities reported here for the NE 
region does not substantially enlarge the existing dataset used by CD and B96,
we do not attempt a complete new kinematic analysis of Coma.  Rather,
we briefly summarize an apparent disrepancy between Paper I, CD, and B96 in
terms of characterizing the kinematics of galaxies in the SW region of Coma.
We then 
describe the main kinematic results found for the other clusters studied by us.  

In Paper I we emphasized that the kinematics of the abnormal spectrum galaxies
in Coma SW are unusual in that the velocity dispersion of the galaxies is
high, $\sim$1400 \kms.  The mean velocity of these galaxies of $\sim$7200 \kms
is also high, when compared to the mean velocity of Coma in the central region.
Using a different method, CD subdivided Coma 
kinematically by fitting two gaussians to the
velocity distribution and find a distinct kinematic substructure in the SW
region.  While their SW kinematic component has a relatively high mean velocity in
the SW region that is similar to that of the abnormal spectrum galaxies, they
find a very low velocity dispersion of 329 \kms for their kinematic substructure.
In fact, according to their analysis the abnormal spectrum galaxies are not
associated with the kinematic substructure.  B96 also analyzed the kinematics
of the abnormal spectrum galaxies and also came to different conclusions than we
found in Paper I.  Specifically, they use the bi-weight estimator, rather than
the classical rms dispersion, to estimate the velocity dispersion of the 
abnormal spectrum galaxies, and find a dispersion of $\sim$300 \kms, but 
with a large uncertainty.  Thus the chief discrepancy centers on how the
velocity dispersion of galaxies in the SW region of Coma, and specifically
the abnormal spectrum galaxies, compares with the rest of Coma.
This subject is re-visited in \S 6, but it should be clear already that it
is critical how one defines the ``velocity dispersion'' parameter in cases where
the velocity distribution may be distinctly non-gaussian.  Thus while there
is agreement that a cluster-cluster merger is seen, there is disagreement about
whether the abnormal spectrum galaxies are involved.

Turning now to the other clusters, we begin with DC0326--53/DC0329--52, which is
probably the most complex of the nearby
rich clusters studied by us, from the kinematic point of view.  
Given that this double cluster can
be conveniently divided at a declination 25\arcmin \ south of the center
of DC0329--52, we have plotted the
observed galaxies in a Dec versus radial velocity plot in Fig. 18.
The plot illustrates several important kinematic features.  First, with the exception
of a single outlier galaxy, the kinematics of DC0326--53, the southern
component in Fig. 18, exhibits a 
velocity gradient as one moves from N to S in the cluster, with the gradient
in velocity amounting to $\sim$500 \kms across the subcluster.  Since there
is no reason a priori why the velocity gradient should be exactly aligned
along the N-S axis, we have made a linear least squares solution for the
velocity gradient projected onto an axis that we have varied in increments of
10\arcdeg .  For each projection we made a linear least squares fit to the
radial velocity versus position along the axis.  A minimum is found in the
reduced $\chi$$^2$ of the fit at a position angle of 55\arcdeg \ North of East, i.e.,
approximately the same position angle that connects DC0326--53 with DC0329--52.
The velocity gradient is near its maximum value at this PA, and corresponds to
a gradient of 18 \kms \ arcmin$^{-1}$, or a total gradient of $\sim$700 \kms.
The significance of this gradient is at the 2.5\% probability level.

Second, the northern component, DC0329--52, in Fig. 18 appears to contain a distinct
group of 9 high-velocity galaxies, which have velocities in the range 18500
\kms \ to 20000 \kms \ and are located from --15\arcmin \ to 0\arcmin \ in
Dec offset from the center of DC0329--52.  While the evidence for such a 
distinct high velocity group is not overwhelming, we also note that 
the rms line-of-sight velocity dispersion (corrected for the relativistic
effect) for all 42 galaxies observed in DC0329--52 is 960 \kms , as opposed to
only 645 \kms \ if the 9 high-velocity group members are excluded.
The high dispersion of 960 \kms , although not unknown for nearby rich
clusters, would place DC0329--52 among the highest 10\% in velocity
dispersion of the 80 rich clusters studied in the ESO Nearby Abell
Cluster Survey  (Mazure et al. 1996), while the 645 \kms \ figure would
make DC0329--52 much more typical of nearby clusters.
Further evidence for substructure
in DC0329--52 can be found in an archival pointed ROSAT PSPC image of the
cluster obtained by Kriss et al. in 1992.  As shown in Fig. 19, where the ROSAT
contours are overlaid on a position plot of the galaxies, there is a clear
double structure to the x-ray emission.  Moreover, the high velocity clump
galaxies form a fairly well-defined swath running from NW to SE, as can be 
seen in Fig. 20,
where the high velocity clump galaxies are marked as filled circles.  The swath
of high velocity galaxies runs across the primary x-ray peak in DC0329--52, 
which has been marked as a large plus in Fig. 20.  In short, the available data
indicates that DC0329--52, a member of a double cluster,  itself consists of 
two kinematic and spatial components.

If we now exclude the high-velocity clump from DC0329--52, there appears to be
a pronounced radial velocity gradient in the remaining 33 cluster members, as can
be seen in Fig. 18.  After projecting the velocity gradient onto an axis that
we varied in 10\arcdeg \ increments (see above) and then making a linear least 
squares fit to velocity versus projected distance, we find that the maximum
reduced $\chi$$^2$ of the fit occurs at a position angle of $\sim$70\arcdeg \ N
of E, and that the gradient is $\sim$52 \kms \ arcmin$^{-1}$ , for a total gradient
of $\sim$1500 \kms \ across DC0329--52.  The mean cz of the 33 galaxies is 17390
\kms \ and the velocity dispersion (corrected for the relativistic effect) is
645 \kms .  The mean cz of the 9 high velocity clump members is 19310 \kms \ 
and their dispersion (again corrected for the relativistic effect) is 480 \kms .
Velocity gradients can be induced by tides during a merger, as we discuss in \S 6.

Finally, we consider whether the abnormal spectrum galaxies have different
kinematics from the normal spectrum galaxies.  Given the complexity of the
DC0326--53/0329--52 system, computing an overall velocity dispersion for the
abnormal spectrum galaxies is not particularly informative.  There are two
abnormals in the proposed high velocity clump in DC0329--52, six more in the
main DC0329--52 cluster, and four in DC0326--53.  As can be seen in Fig. 18,
there is no evidence that the kinematics of the normal and abnormal
spectrum galaxies are different within these substructures.

Turning now to DC2048--52, the Dec offset versus redshift data is illustrated
in Fig. 21.
If this cluster is divided into a main cluster, containing all galaxies with Dec
offset less than 30\arcmin \ north of --52:59:01 (an obvious dividing line), 
and a northern subcluster, containing
all galaxies greater than 30\arcmin \ north, then there are 22 galaxies in the
northern subcluster that we have observed.  A search for velocity gradients
in the northern group was made, after restricting the sample to only
those galaxies that are not definitely abnormal-spectrum, throwing out one 
galaxy with high velocity (17000 \kms) and one with low (9000 \kms).  Only a 
very weak case for a velocity gradient (probability for a random sample $\sim$11\%)
can be made for this restricted sample of 14 galaxies.  The gradient, 
if real, is
approximately along the line of centers joining the northern component with the
main cluster.

If we now restrict ourselves to only
the 6 abnormal-spectrum galaxies, we find a somewhat more
convincing correlation of velocity and position 
(probability only $\sim$3\% for a random sample), along the same axis as 
the normal galaxies but with a velocity
gradient {\it opposite} in direction and about twice as large.
While we find modest evidence in favor of a
radial velocity gradient in both the normal and abnormal spectrum galaxies
in the northern component,
clearly, the evidence is not nearly as strong as in the case of DC0326/0329. 

A glance at Fig. 21 reveals that there is a substantial difference
in mean velocity between the main cluster and the northern component, 
as originally shown by Dressler \& Schectman(1988b).
Specifically,
the mean cz for the 61 galaxies in the main cluster is 13419 \kms \
and the velocity dispersion  is
788 \kms .  The mean cz of 14360 \kms \ for the northern component is nearly 
1000 \kms \ higher than the main cluster, while the
velocity dispersion  is 
441 \kms .  There is weak evidence that the abnormal spectrum galaxies
in the main cluster have a higher velocity dispersion than do the
normal galaxies (992 versus 732 \kms).
The main cluster exhibits no indication of a systematic velocity
gradient.

DC2048--52 has been observed with the ROSAT PSPC in pointed mode.
Unfortunately, that image is centered between the main cluster and the
northern subcluster, so that the main cluster is bisected by the 
vignetting from the support structure in the PSPC.  Thus it is 
difficult to derive
solid inferences concerning the nature of the X-ray emission in DC2048--52.
Nevertheless, the image does appear to be asymmetric, with a general
elongation in the N-S direction.

We consider now the double cluster DC0103--47/DC0107--46. 
Our velocity data confirms the main conclusions found by
MKDFR, and adds some extra perspective on the DC0103-47 component.  Basically,
MKDFR found strong evidence for kinematic substructure in the cluster.  The
most dramatic case involves a background group of galaxies at $\sim$9000
\kms , which MKDFR refer to as a ``sheet'', owing to the spatially extended
nature of this group and to the narrow velocity range near 9000 \kms \ that
it occupies.  Evidence for the sheet is presented in Figs. 4, 5, and 7 of
MKDFR.  Our kinematic data, which is less extensive than MKDFR's, does not
show the sheet as convincingly as MKDFR's wedge diagram in Fig. 4 of their
paper.  However, in Fig. 22 of this paper one can see evidence of the sheet in
the southern (DC0103--47) part of the double cluster, while in the main 
cluster there is no such well-defined demarcation between cluster and sheet.
It is interesting to note that 4 of the 6 abnormal spectrum galaxies found by
us have velocities at or close to the sheet.  The other 2 abnormals have 
very low velocities characteristic of the ``Group B'' found by MKDFR as a
kinematically distinct component.  Thus, although the small number statistics
preclude a definite conclusion, the abnormal spectrum galaxies appear
to lie on the margins of the velocity distribution, an effect that has also been
noted for the blue galaxies in distant clusters (Dressler 1987). We also note the
presence in Fig. 22 of the ``Group C'' component found by MKDFR, specifically,
a low velocity component in the SW region (i.e., in DC0103--47).  In fact, of
the 12 galaxies more than 20 \arcmin \ South of the center of DC0107--46
that are cluster members,
7 are members of the low velocity component, 4 have velocities near that of
the sheet, and a fifth also has fairly high velocity.  The overall picture
emerging from MKDFR's and our study is that the DC0107--46 cluster is
located  between the high velocity sheet and the low velocity SW subcluster.
In fact, it may be acquiring galaxies through infall from either the high
or low velocity groups.

There is also a pointed ROSAT PSPC image of DC0107--46 in the NASA public 
archive.  The X-ray image (not reproduced here) shows a centrally concentrated emission in the
center of DC0107--46, and a fan-shaped asymmetric emission pattern that points
along the N-S axis.  There are numerous point sources to the South of the
concentrated emission.

Finally, we briefly consider the cluster DC1842--63, which was also 
studied extensively by MKDFR.  The latter authors found no evidence for
spatial or kinematic substructure in DC1842--63.  With our very limited
data set for this cluster, we have little to add to the analysis of MKDFR.
The one potentially interesting point is that the pronounced starburst
galaxy found by us (with a D value of 7.6) also has the lowest velocity.
Thus again we see some evidence that the abnormal spectrum galaxies tend
to lie on the margins of the velocity distribution.

\section{Discussion}

The principal conclusions to emerge from the data presented in the previous
section are twofold.  First, abnormal spectrum early-type galaxies are generally
present  in nearby rich clusters, and many of
these galaxies have spectral signatures that are reminiscent of the starburst 
and post-starburst activity seen prominently in distant clusters.  Given
the higher abundance of PSB galaxies in our present-epoch rich cluster sample  
than in Zabludoff et al.'s field survey of nearby galaxies, it thus appears that the rich
cluster environment is still an ongoing catalyst for galaxy evolution of the
kind rarely seen in the field.  Second, with the exception of DC1842--63,
there is evidence in each of the clusters that dynamical equilibrium has not
yet been established.  This evidence consists of either systematic velocity
gradients and/or velocity differences between different components.  We
begin by considering the second issue of velocity substructure, specifically,
the origin of the systematic velocity gradients seen in DC0326--53/0329--52,
in Coma, and perhaps in DC2048--52.

A velocity gradient can be due to either rotation, a non-radial shear, or a
radial distension.  In all three cases the tidal action of one structure on
another provides a logical source for the gradient.  To assess this
possibility, we have made use of the publicly available NEMO stellar
dynamical toolbox (Teuben 1994) to simulate the tidal field of a rich cluster
on an infalling subcluster.  In most cases our initial conditions consisted of
a main cluster with 512 particles and a subcluster with 128 particles, each 
an equilibrium Plummer model.  A softening parameter of 0.05
mean harmonic radii was used in the particle potentials to account for the
extended mass distributions in galaxies. The two structures were 
typically placed 4 mean harmonic radii apart, and then a variety of initial 
purely radial infall velocities for the
subcluster were tried, ranging from parabolic down to one quarter of parabolic.
The simulations were carried out using the hierarchical N-body Treecode 
(Barnes \& Hut 1989) routine in NEMO.  They are very similar in purpose to the
numerical simulations of infalling subclusters reported by Roettiger et al. 
(1993), Burns et al. (1994a), and Burns et al. (1994b), except that they do
not follow the hydrodynamics of the ICM as the latter authors have done.
In all cases, a strong velocity gradient is produced by
tidal distension of the infalling subcluster, {\it after} the subcluster has 
passed through the main cluster and is on its way out again.  The production
of a gradient in the subcluster after passage through the main cluster is
consistent with the Burns et al. (1994a) simulations of the Coma 
cluster, in which they found that the strong dynamical effects become apparent
after the subcluster has fallen through the main cluster.  In addition, 
after passing through the main cluster the subcluster loses its central 
concentration and becomes widely dispersed, which again is consistent with
what Roettiger et al. (1993) and Burns et al. (1994a) report from their 
simulation.  Not surprisingly, with
a four to one mass ratio between cluster and subcluster, there is essentially 
no velocity gradient produced in the main cluster as a result of the encounter.

The results of one particular simulation, in which the subcluster fell in with 
25\% of the parabolic velocity, are shown in Figs. 23 and 24.  In Fig. 23, which
is a two-dimensional X-Y position plot (where the X axis defines the infall
direction of the subcluster and direction of initial separation between
cluster and subcluster), the
four panels show the main cluster and subcluster at four different time steps.
In Fig. 23(a) the simulation begins with the cluster and subcluster separated by
four mean harmonic radii.  In Fig. 24, which shows an X versus V$_X$ 
position-velocity plot, the development of the velocity gradient after the
subcluster passes through the main cluster can  be seen.  Note that as a 
result of the encounter, $\sim$60\% of the subcluster particles have joined
the main cluster by the final timestep shown in Fig. 24(d); Roettiger et
al. (1993) found a similar cannibalization of the subcluster in their
combined gas dynamical/N-body simulation of an infalling subcluster. Other purely
radial simulations have been tried, with higher (up to parabolic) infall
velocities, and produce very similar results.  However, for the higher infall
velocities, there is a large (and obvious) velocity offset between the 
subcluster and the main cluster, and a much smaller fraction of the subcluster
is assimilated by the main cluster during the encounter.  We also ran a 
non-radial infall simulation in which the sub-cluster was offset from the
initial velocity vector by one mean harmonic radius.  The results were 
essentially the same as for the purely radial infall cases.

To assess whether tidal torquing of a non-spherical infalling subcluster could
produce a significant rotational gradient, we also set up a simulation in which
the subcluster was composed of two Plummer models separated by one harmonic 
radius in both the X and Y directions (where again the X axis defines the 
initial infall direction and line of separation between subcluster and main
cluster).  Thus we have roughly simulated a prolate subcluster
oriented at a 45\arcdeg \ angle to the main cluster.  To give the tidal torque a 
greater ability to operate, we started the simulation with the prolate 
subcluster at a distance of 12 mean harmonic radii.  The main result is that
no strong rotational gradient is produced on the way in.  In short, our
simulations indicate that a velocity gradient arising from a radial tidal
distension can be readily produced during the passage of a subcluster through
the main cluster, but that rotation or non-radial shear are not easily
produced and, if present, must be a result of the ``initial conditions''.

We now compare the results of the simulations to the specific cases of the
clusters we have studied. To begin with, the velocity gradient in DC0326--53,
as well as the generally dispersed nature of the subcluster, appears to be
readily explained through tidal distension of DC0326 by DC0329.  If so, we must
conclude that DC0326 has already passed through DC0329.  This inference is
both interesting and perhaps surprising, since the known abnormal spectrum galaxies
in both DC0326--53 and DC0329--52 are mostly of the emission line
variety, i.e., they are either currently starbursting and/or AGN.  The
implication is that the starbursts are provoked only after the subcluster has
passed through the main cluster.  In Coma, where Burns et al. (1994) have
strongly advocated that the SW subcluster has also passed through the main
cluster, and may even in fact be falling back in for the second time, 
they have estimated that the SW structure passed through the
main cluster $\sim$2 Gyr ago.  Caldwell et al. (1996) have argued on the basis
of spectral indicators that the PSB galaxies in the SW are typically $\sim$1 Gyr
post-starburst.  Thus in both Coma and DC0326--53/0329--52 the inference is that 
the starbursts occurred only after the subcluster had passed well through the
main cluster, which provides an important constraint on the mechanism 
responsible for the starbursts (see below).  It should be noted that since our
simulations indicate that a large fraction of the subcluster is cannibalized by
the main cluster during the tidal passage, many of the abnormal spectrum
galaxies in DC0329--52 and in the central region of Coma may actually have
originated in the subclusters.

While the spatial and kinematic structure of DC0326--53 appears to be the
result of tidal distension, that of DC0329--52 itself remains difficult to
interpret, since, as mentioned before, our simulations do not produce a strong velocity
gradient in the main cluster, unless the cluster/subcluster mass ratio is
nearly unity.  Thus the large apparent velocity gradient in the main part of
DC0329--52 (i.e., excluding the high velocity clump) is puzzling.  However,
the clear presence of substructure in the main cluster, as evidenced by the
two x-ray peaks and the high velocity clump, indicates that the main cluster
is itself undergoing strong dynamical evolution at the present time, and thus
is currently in a complex kinematic state.  A potentially important constraint
is that the position angle of the velocity gradient in the DC0329--52 main 
cluster is nearly coincident with that of the DC0326--53 subcluster.  

The case of the DC0103--47 subcluster, and its relation to the DC0107--46 main
cluster, is somewhat less clear than that of the DC0326--53 subcluster.  The
rather sparse data on DC0103--47 show no evidence for a velocity gradient.
Thus a tidal signature is not immediately obvious.  Instead, the main 
kinematic signatures, as discussed previously, are the presence of the high
velocity background sheet and the low velocity of the the remaining galaxies 
in DC0103--47 (relative to the main DC0107--46 cluster).  However, the
DC0103--47 subcluster has a very strongly dispersed appearance in comparison to
the main cluster.  This latter fact leads us to suspect that DC0103-47 has in
fact already passed through DC0107--46, although the evidence is not as 
strong as in the case of DC0326--53.

The situation regarding DC2048--52, and its northern subcluster, is even less
clear.  There is weak evidence for a velocity gradient in the northern
subcluster, both in the normal and abnormal spectrum galaxies, but the 
evidence is far from compelling.  In addition, the northern subcluster is
not as centrally concentrated as DC2048--52, but neither is it as dispersed
in appearance as DC0103-47 or DC0326--53.  Thus we can say little about the
current dynamical condition of the cluster.

As mentioned earlier, the dynamical state of Coma has been extensively discussed by CD and by B96.
We have little to add other than to point out that we have now demonstrated
that the fraction of abnormal spectrum galaxies is substantially higher in
the SW region of the cluster than in the center or NE.  Hence it now appears
highly likely that the abnormal spectrum galaxies in the SW are indeed 
connected with the x-ray and optical substructure observed in that area.  
In short, a connection must be sought between the merging subcluster and
the triggering of starbursts in galaxies in that region.  The available 
evidence, compiled by Burns et al. (1994a), indicates that the SW substructure
is emerging from having fallen through the main cluster, or perhaps even 
infalling for the second time.  The velocity gradient in the central region
of Coma along the NE-SW axis found by both CD and B96 is certainly
difficult to explain on the basis of tidal action, given the large mass of
the main cluster.  However, the complex situation regarding the kinematics
of NGC 4874 and NGC 4889, whose velocities are reversed from the general
gradient (as is the velocity of the SW substructure; cf. B96), allows for
a variety of infall situations involving multiple subclusters.  

There are several things to keep in mind in this discussion:  \\
(1) Many nearby rich clusters have major subcluster structures.  This is 
obvious from a glance at the pronounced double structures of some of the
clusters, and the point has been convincingly made before (e.g., Geller \&
Beers 1982; Fitchett \& Webster 1987; Dressler \& Shectman 1988b; West \& 
Bothun 1990; Bird 1994).\\
(2) From a variety of arguments, Burns et al. (1994a) have proposed that the
substructure now seen in the SW of Coma has already passed through the main
body of Coma and is now re-emerging (or even infalling for the
second time).  In some, if not most, of the rich clusters studied by us 
(including Coma), we also find evidence that the subcluster
has passed through the main cluster and out the other side.  The evidence is
based on the presence of velocity gradients and/or 
the dispersed appearance of the subcluster.  \\
(3) Abnormal spectrum galaxies are common in nearly all of the rich clusters that we
have studied.  Based on the currently starbursting or recent post-starburst
spectroscopic signatures of many of these galaxies, coupled with the above
inference that the subclusters in at least some cases have passed through
the main cluster, it appears that much of the starburst activity has been 
triggered only when the subcluster passes through the main cluster or perhaps 
even afterwards.

We now seek a connection between the infall of subclusters and the late 
triggering of starbursts in individual galaxies.  While several possibilities
exist, we propose that strong shocks in the intracluster medium generated by
the passage of a subcluster is perhaps the most plausible triggering mechanism
for the late starbursts.  Burns et al. (1994a,b) have shown that in their 
infalling subcluster hydrodynamical/N-body simulations, in which the 
cluster/subcluster ICM's experience a
strong collision, extensive intracluster shocks are produced (given that
infall velocities are well above the local sound speed) that propagate 
through the ICM for several Gyr.  These shocks could account for the
correlation between the presence of wide-angle tailed (WAT) radio sources, and
the presence of x-ray substructure in clusters found by Burns et al. (1994b).  
In addition, Burns et al. (1994b) propose that the bending of
many WAT sources is produced by subcluster-induced 
bulk motions in the ICM sweeping past a radio source, rather than intrinsic
motion of the galaxy harboring the radio source (this is a particularly
appealing explanation for WAT's associated with slowly-moving cD galaxies).  
We propose that the Burns et al. shocks may also be responsible for 
triggering the starbursts in the numerous SB/PSB galaxies that we have found
in nearby clusters.  For this mechanism to work, the ICM shocks must be
capable of sufficiently compressing molecular clouds in galaxies to trigger
star formation.  Thus, the progenitor galaxies must implicitly have had
a substantial ISM prior to entering the cluster, and were therefore probably
of a later Hubble type than we now see.
Since it has been proposed before that ram pressure from a
hot ICM should trigger star formation in infalling cluster galaxies (e.g., 
Dressler 1987; Gunn 1989),
it seems plausible that intracluster shocks could produce the same effect.

\section{Conclusions}

In the Introduction, several questions were posed regarding the apparent
absence of SB/PSB galaxies in present-epoch rich clusters.  Based on the
results presented here, it appears that the star formation observable in
such a large fraction of cluster galaxies at redshifts of z$\sim$0.3-0.5
still occurs at the present epoch, though at a much lower level (3-12\% for
PSB galaxies depending on the exact definition employed).
Perhaps the most significant difference
between z$\sim$0.3-0.5 and the present epoch is that the starburst signature
is generally stronger in the individual distant cluster galaxies.  The
lower starburst strength for nearby galaxies 
may be a consequence of the strong depletion
in HI that is observed for the outer parts of 
galaxies in rich clusters (e.g., Gavazzi
1989),
thus leaving a centrally concentrated HI distribution (Cayatte et al. 1994).
It is thus not surprising that Caldwell et al. (1996) found that the
starbursts which occured in the Coma cluster SB/PSB galaxies were  
centrally concentrated.  Similarly, Moss and Whittle (1993) have found 
that cluster spirals with strong H$\alpha$ emission tend to be {\it early-type}
spirals with compact nuclear emission, rather than disk-wide emission.  
So over time, the amount of material in galaxies that is available for
starbursts has declined and become more centrally concentrated.

The answer to the third question posed in the Introduction, concerning the 
precise mechanism responsible for the starbursts in cluster galaxies, remains 
highly uncertain.  While it is curious that
some PSB galaxies had an origin unrelated to clusters (Zabludoff et al. 1996),
it seems to be the case that clusters are the preferred environments for 
inducing starbursts at low redshift, and perhaps at high redshift as well.
We have speculated that shocks in the ICM produced by
the infall of a subcluster could trigger the starbursts, but other
mechanisms may be also valid, such as the ``galaxy harrassment'' scenario 
suggested by Moore et 
al. (1996), in which tides raised during high speed encounters between
cluster galaxies cause resident gas to lose angular momentum, sink to the
galactic centers, and create starbursts.  This method may have some difficulty
in explaining the particular case of Coma, since most of the Coma PSB
galaxies are in a restricted region and have similar starburst ages
(Caldwell et al. 1996), but it may well be appropriate for the other clusters.

To summarize, our study indicates that there is a likely connection between
the presence of SB/PSB early-type galaxies in clusters and the existence of
substructure in those clusters.  Furthermore, evidence from observed
velocity gradients and/or the dispersed appearance of the subclusters 
indicates that in some of the cases the subcluster has already passed through
the main cluster.  The relative youth of the SB/PSB phenomenon in the
Coma cluster (less than 1 Gyr old) and in DC0326--53/0329--52 (most of the
abnormal spectrum galaxies still in a current star formation state) then
implies that the triggering of the starbursts occurred only after the
subcluster passed through the main cluster.

Since DC0326--53/0329--52 provides the clearest case of a subcluster that has
already passed through the main cluster (viz., the velocity gradient in
the DC0326--53 subcluster), and since the majority of the abnormal spectrum
galaxies are currently starbursting, rather than PSB, the DC0326--53/0329--52
system probably offers the most promising case for further understanding the
connection between substructure in clusters and the presence of SB/PSB
galaxies.  Since we have acquired spectra for only $\sim$40\% of the
galaxies in Dressler's (1980) list, a further multi-fiber spectroscopic study
should be highly useful in improving the definition of kinematic substructure
as well as the frequency and spatial location of the abnormal spectrum galaxies.
In addition, high-resolution imaging of the currently starbursting galaxies
could provide much additional insight concerning the mechanism that has
triggered the starbursts. 

\acknowledgements

Some of the observations reported here were obtained with
the Multiple Mirror Telescope, which is jointly run by the
Smithsonian Institution and the University of Arizona.
We thank Richard Bower, Richard Ellis, and Ray Sharples for permission to use
our jointly acquired Coma cluster data for this paper.  Thanks are also due to
Eliot Malumuth for providing coordinates for DC1842--63 and DC0107--46, to Jon Morse
for providing long-slit spectra of two galaxies in DC0326--53/DC0329--52, and to Rob
Kennicutt for making available to us his atlas of global spectra of galaxies.
Suzanne Tourtellotte mercifully measured the standard stars on the CTIO 0.9-m
telescope images.  We also thank Peter Teuben for invaluable help in the 
installation of
the NEMO softeware and for advising us in its use.  Finally, we acknowledge
helpful discussions with Hans Eberling and Alistair Edge concerning the X-ray
emission in DC0329--52.

\clearpage

% Now comes the reference list.  In this document, we used \cite to call
% out citations, so we must use \bibitem in the reference list, which
% means we use the LaTeX thebibliography environment.  Please note that
% \begin{thebibliography} is followed by a null argument.  If you forget
% this, mayhem ensues, and LaTeX will say "Perhaps a missing item?" when
% you run it.  Do not call us, do not send mail when this happens.  Put
% the silly {} after the \begin{thebibliography}.
%
% Each reference has a \bibitem command to define the citation format
% and the symbolic tag, as well as a \reference command which sets up
% formatting parameters for the reference list itself.
%
% If we had not bothered with the \cite-\bibitem business, calling out
% the references outselves, the reference list could be enclosed in
% a references environment (\begin{references} has no null argument),
% and there would be no need for the leading \bibitem's.

%\begin{thebibliography}{}

%\end{thebibliography}

% Finally, we have figure captions.  Usually these must be on a separate
% page, although unlike table, it is often permissible to have several
% figure captions on the same page.  We force the page break between
% the reference list and the figure captions.
%
% The \caption command in the figure environment works like the one in the
% table environment (it's the same one, actually), except that this one
% produces identification text that reads "Figure N."

\newpage
\begin{figure}
\plotone{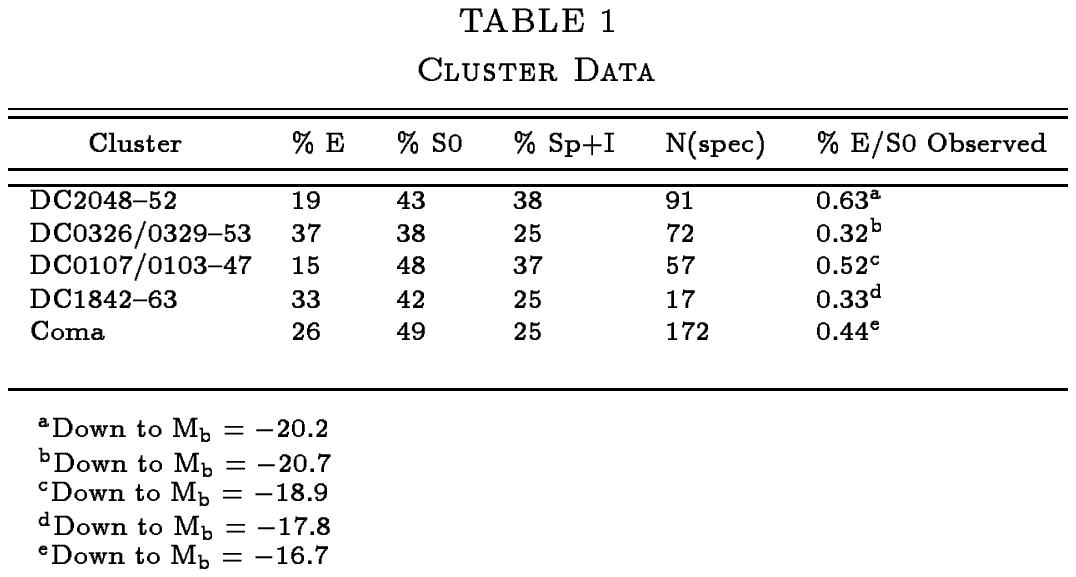}
\end{figure}
\vfill\clearpage
\begin{figure}
\plotone{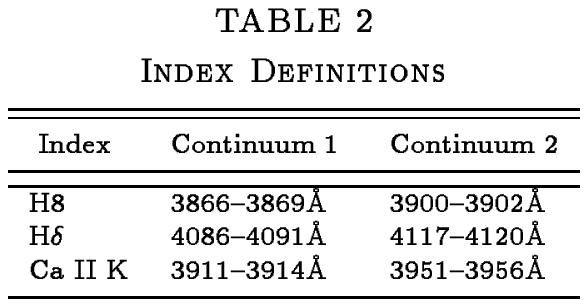}
\end{figure}
\vfill\clearpage
\plotone{caldwell.tab3.ps2}
\vfill\clearpage
\plotone{caldwell.tab3.ps3}
\vfill\clearpage
\plotone{caldwell.tab3.ps4}
\vfill\clearpage
\plotone{caldwell.tab4.ps2}
\vfill\clearpage
\plotone{caldwell.tab4.ps3}
\vfill\clearpage
\plotone{caldwell.tab5.ps2}
\vfill\clearpage
\plotone{caldwell.tab5.ps3}
\vfill\clearpage
\plotone{caldwell.tab6.ps2}
\vfill\clearpage
\plotone{caldwell.tab7.ps2}
\vfill\clearpage
\plotone{caldwell.tab7.ps3}
\vfill\clearpage
\plotone{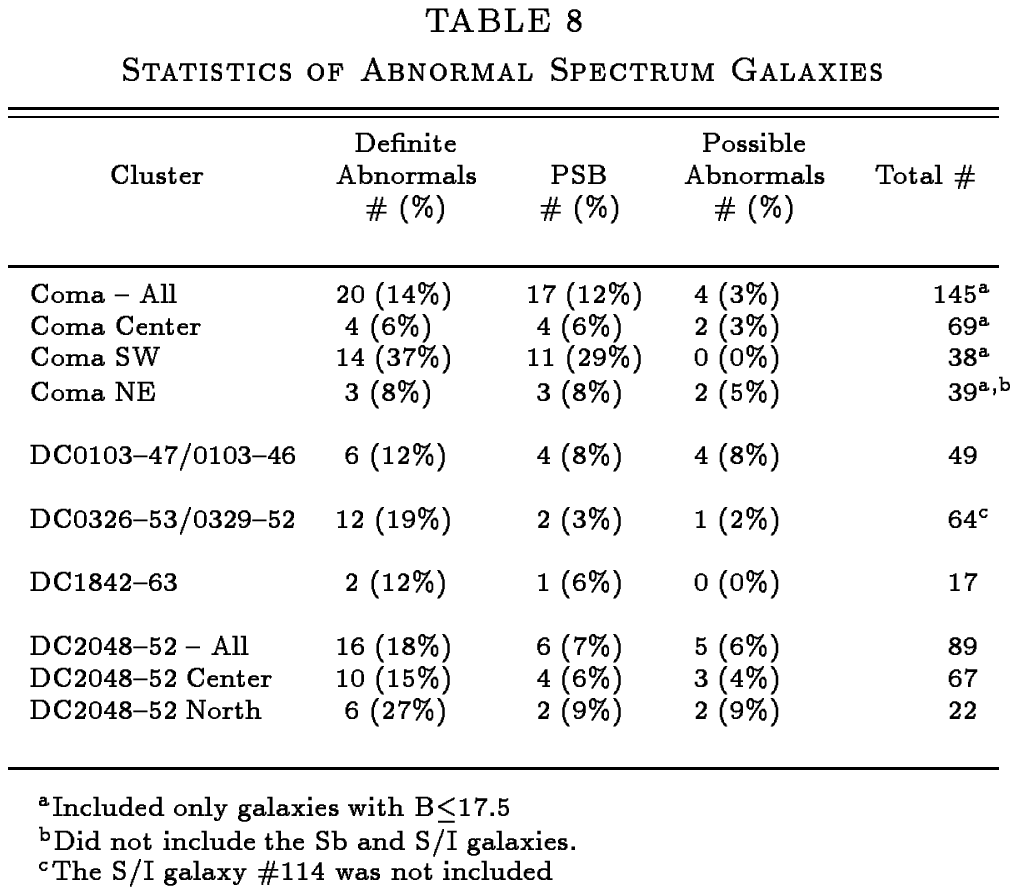}
\vfill\clearpage

\begin{figure}
\figurenum{1}
\plotone{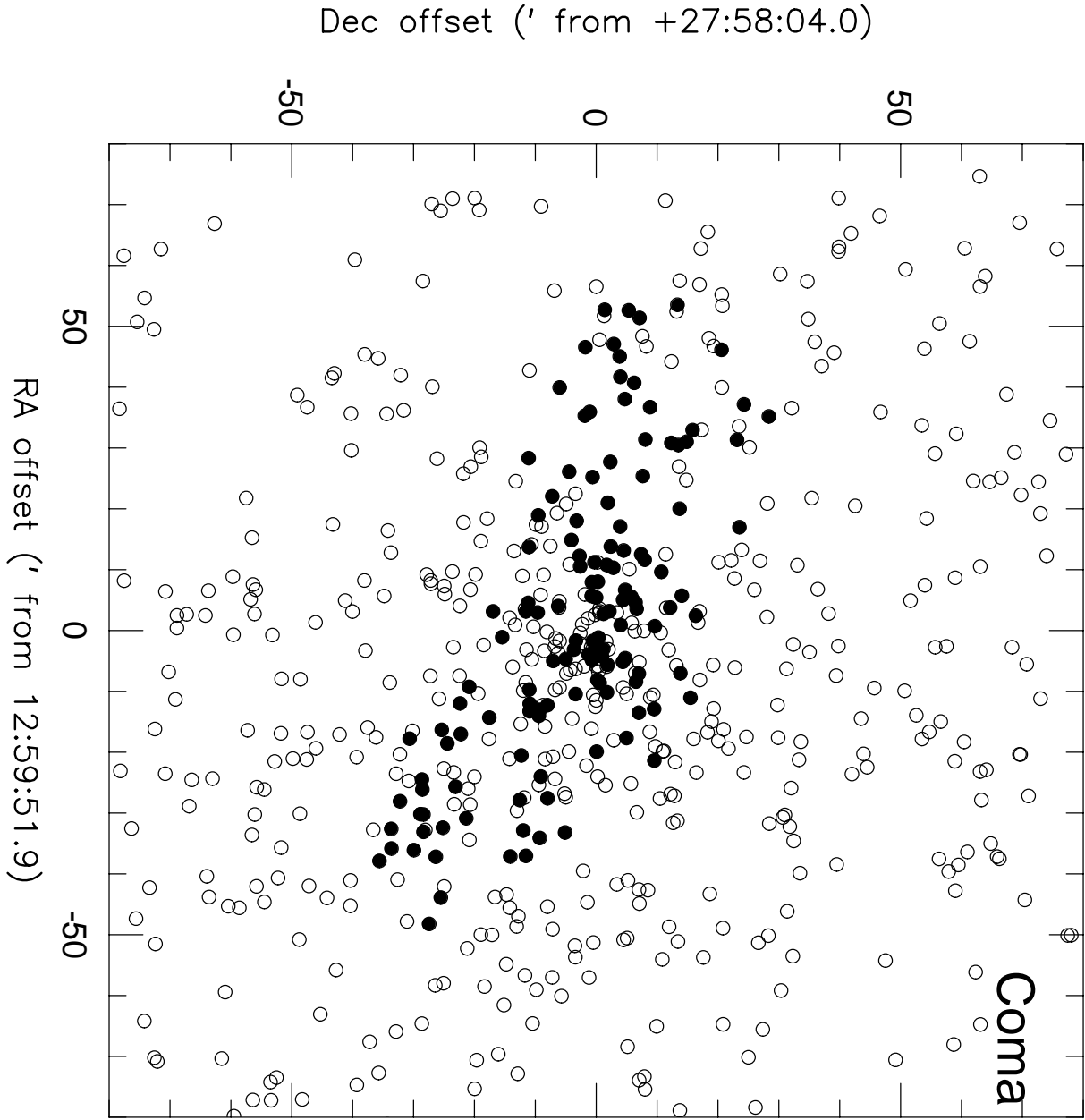}
\caption{Graphical summary of our spectral observations in the Coma cluster.  
Observations reported here or in Caldwell et al. (1993) are plotted as
filled circles; the remaining galaxies in the GMP
catalog with B $\leq$17.5 of all types are plotted as open circles.}
\end{figure}
\vfill\clearpage

\begin{figure}
\figurenum{2}
\plotone{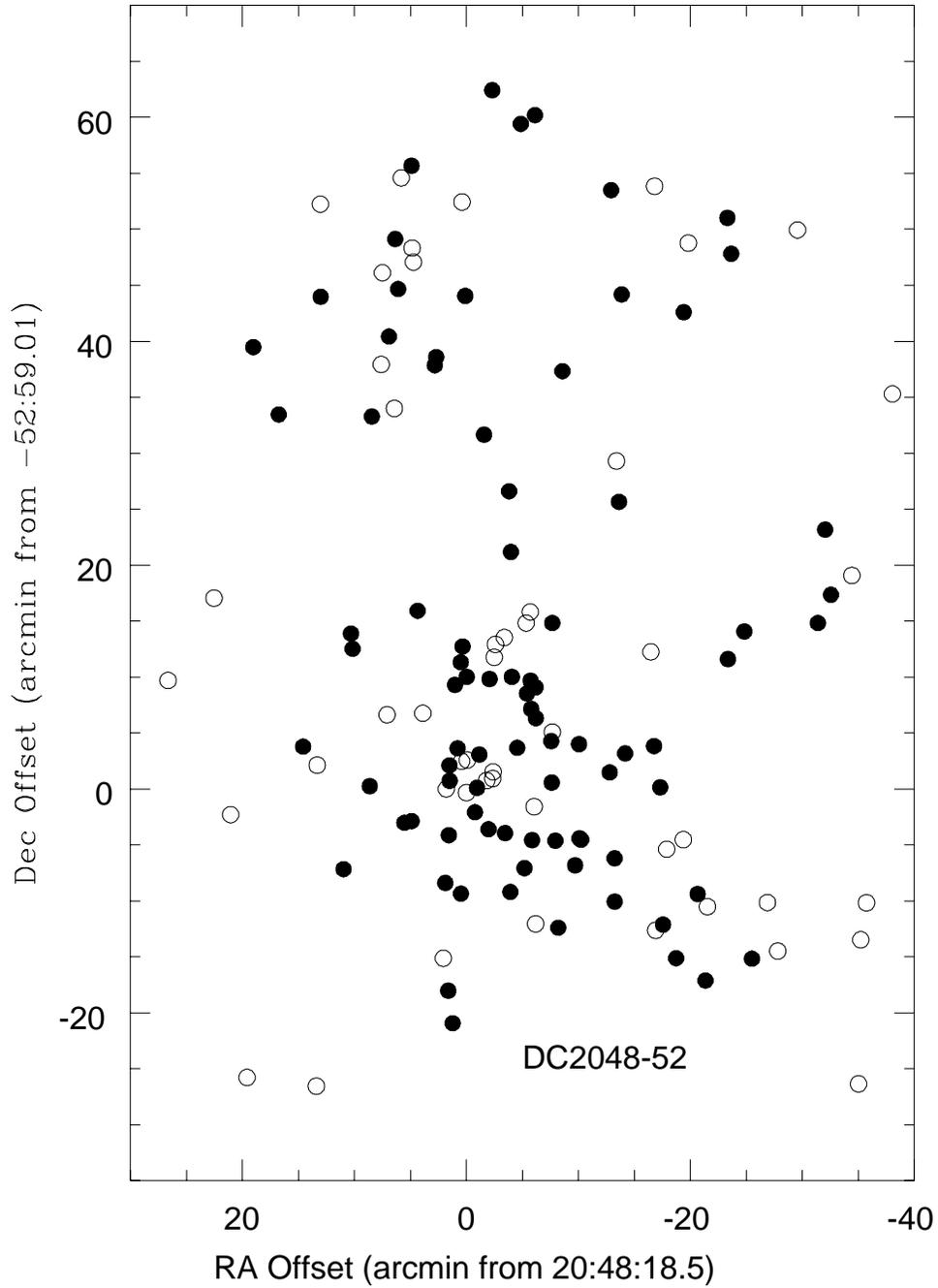}
\caption{Summary of spectral observations in DC2048--52.  Galaxies observed
by us are plotted as filled circles; additional galaxies in
Dressler's (1980) catalog are plotted as open circles.}
\end{figure}
\vfill\clearpage

\begin{figure}
\figurenum{3}
\plotone{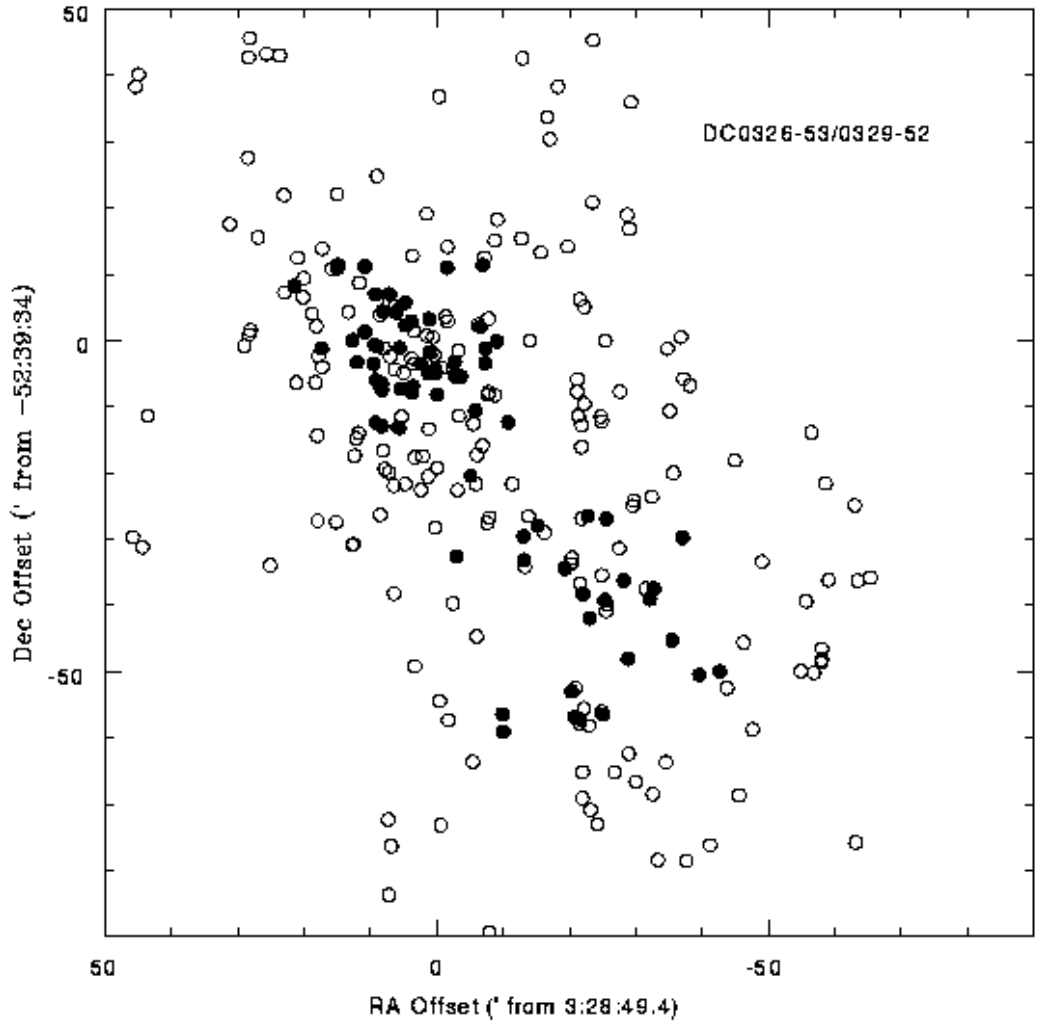}
\caption{Summary of spectral observations in DC0326--53/0329--52.  Same symbols
as in Fig. 2.}
\end{figure}
\vfill\clearpage

\begin{figure}
\figurenum{4}
\plotone{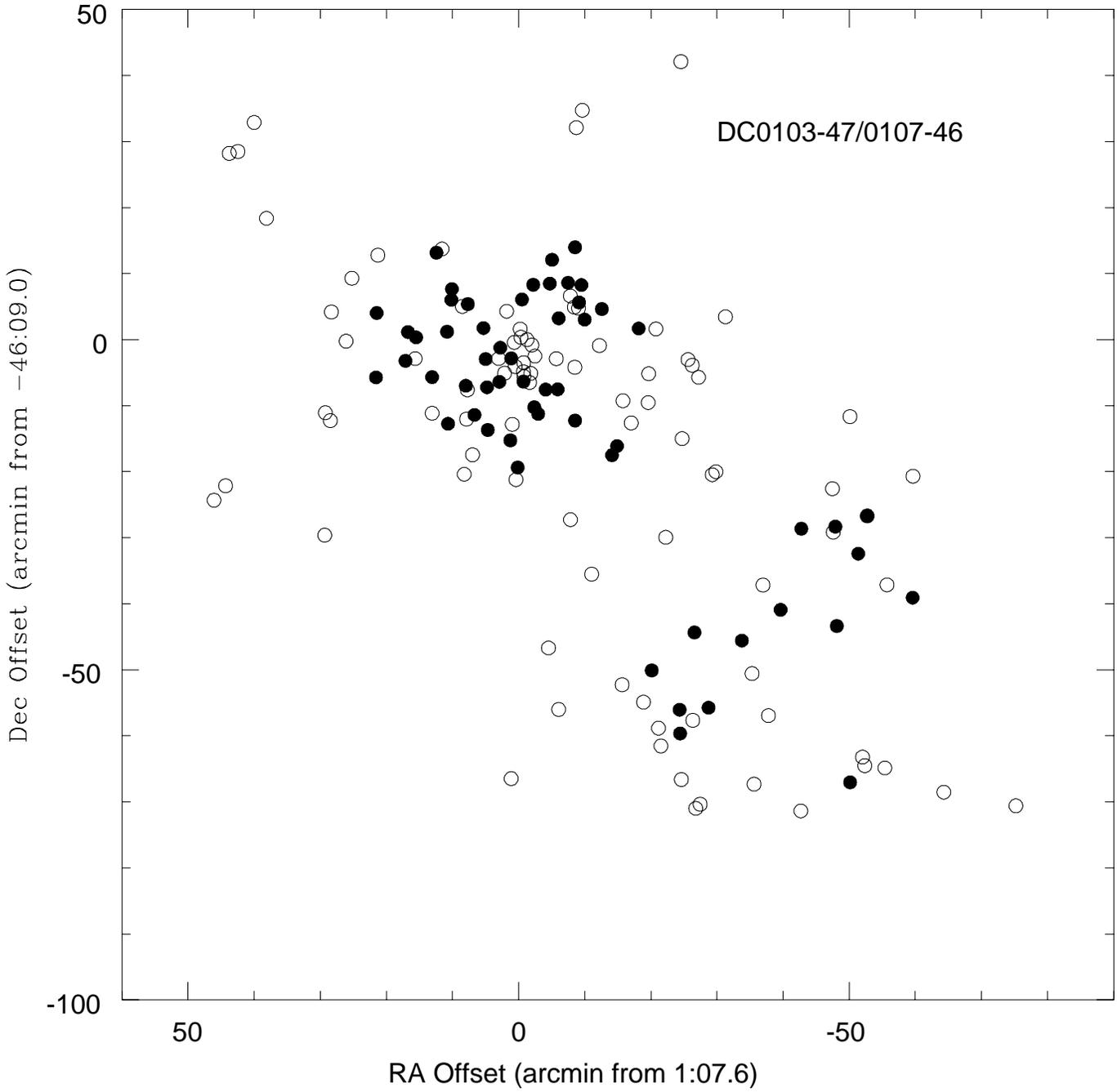}
\caption{Summary of spectral observations in DC0103--47/0103--46. Same symbols as in 
Fig 2.}
\end{figure}
\vfill\clearpage

\begin{figure}
\figurenum{5}
\plotone{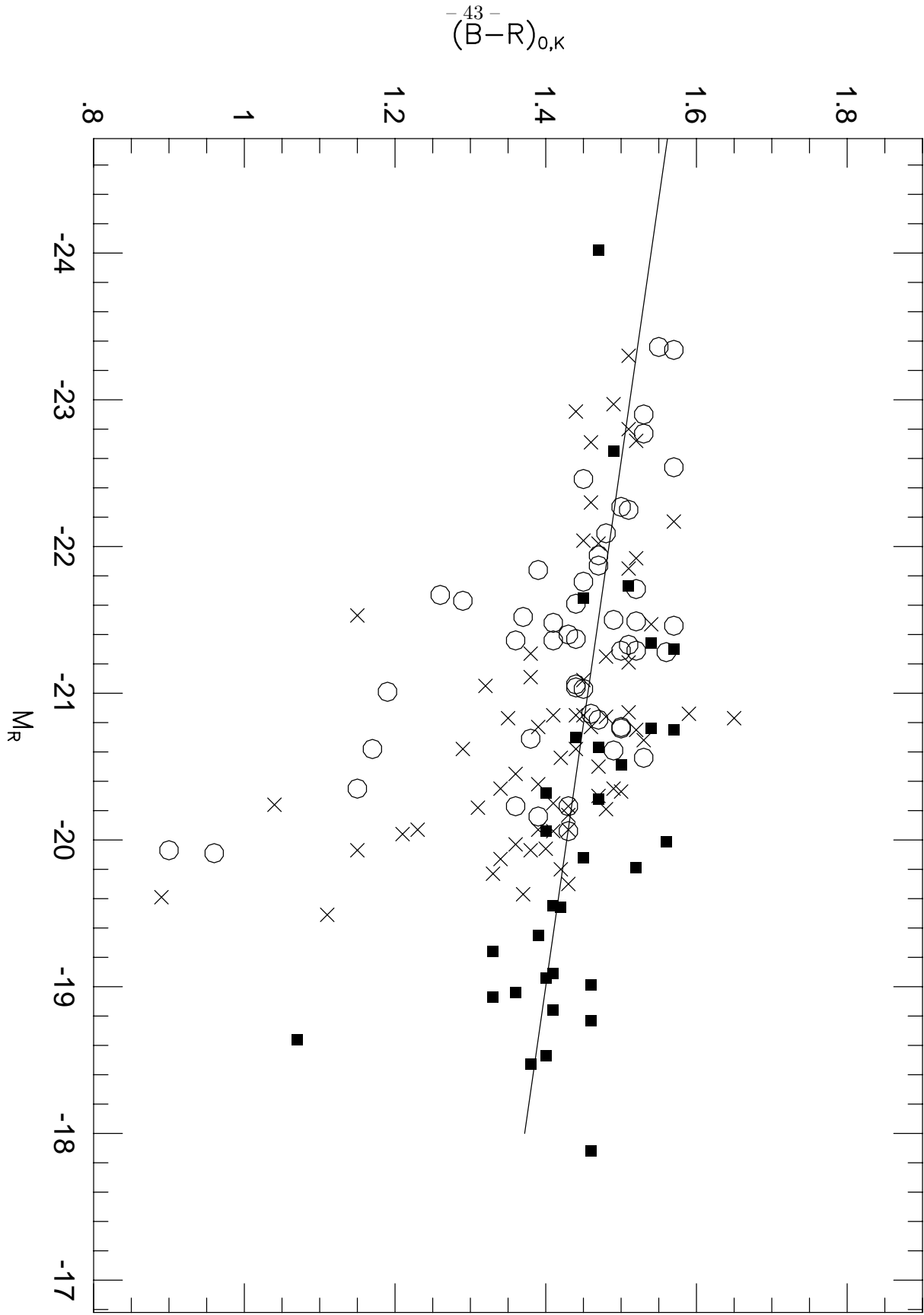}
\caption{Combined c-m relation for DC2048--52 (x's), DC0326--53/0329--52 (open
circles), and DC0103--47/0107--46 (filled squares).  The assumed 
mean relation for normal spectrum galaxies is shown as a line.}
\end{figure}
\vfill\clearpage

\begin{figure}
\figurenum{6}
\plotone{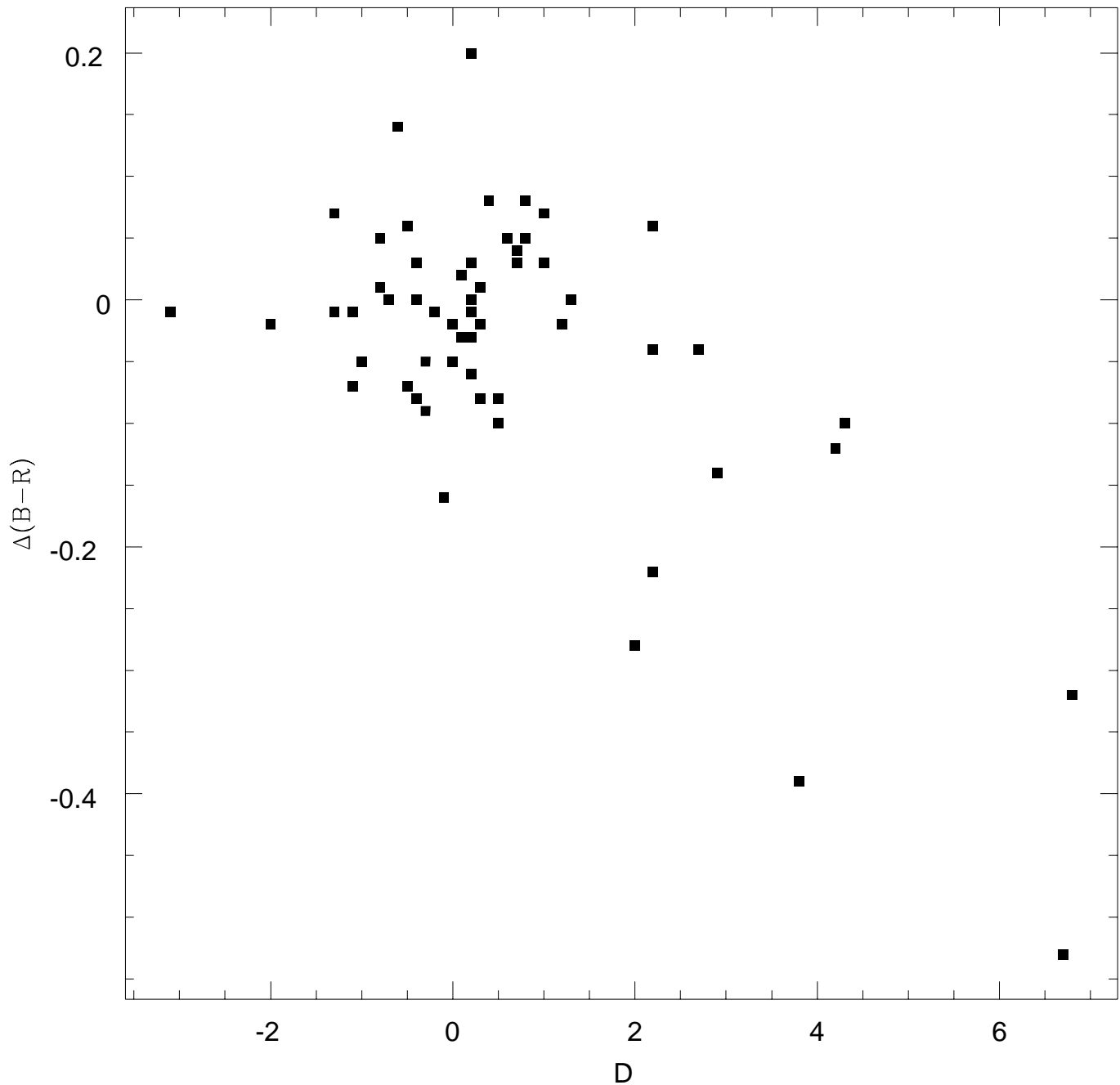}
\caption{D values are plotted against $\Delta$(B--R) color excess for galaxies
in DC2048--52 to illustrate the general correlation between spectroscopic and
photometric measures of abnormality.}
\end{figure}
\vfill\clearpage

\begin{figure}
\figurenum{7}
\plotone{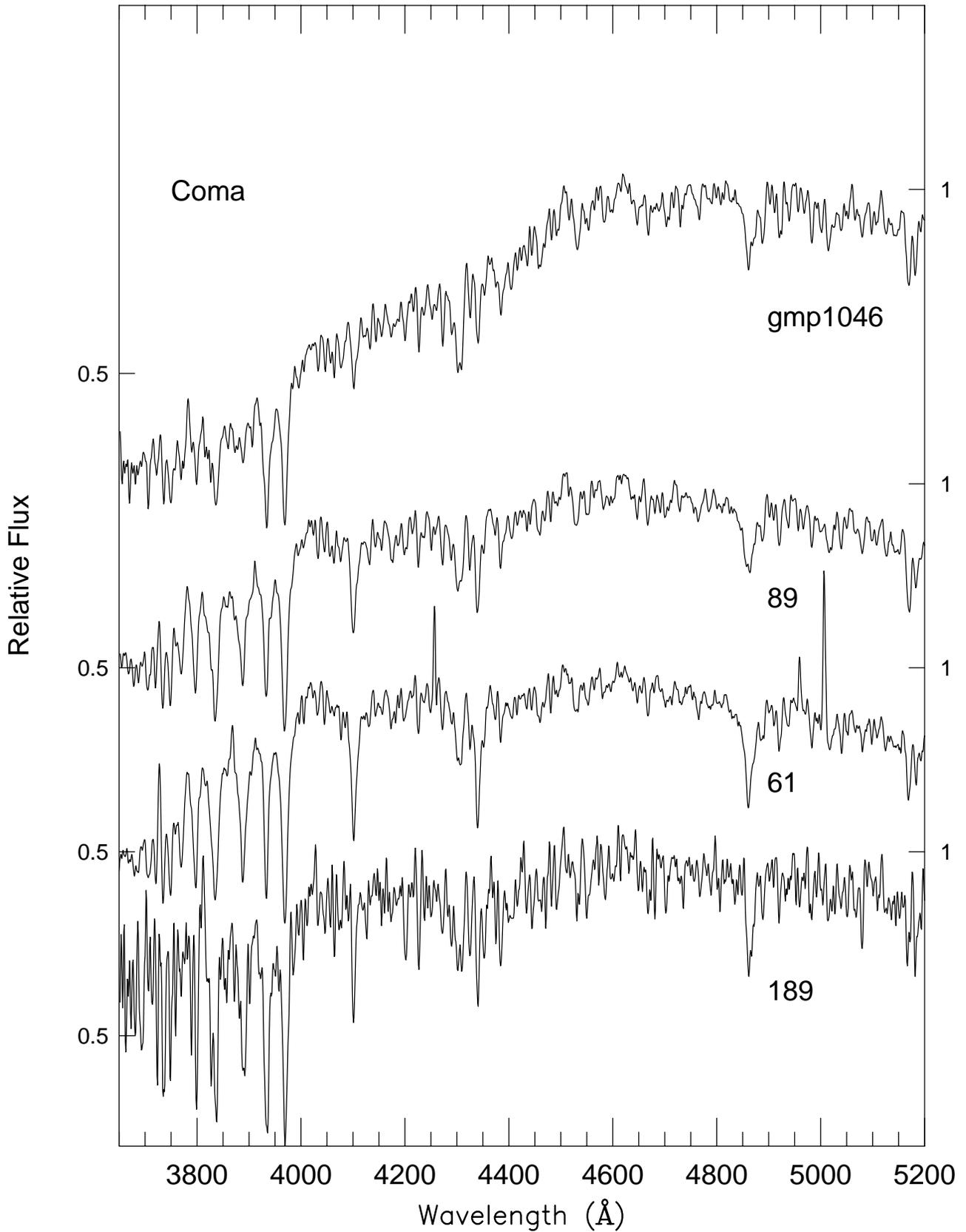}
\caption{Spectra of new PSB galaxies in Coma. All data have been fluxed;
relative flux scales are indicated on the vertical axes. \#189 was mentioned
in Caldwell et al. (1993) as a possible PSB - a better spectrum is shown here
in confirmation of its being called a PSB.}
\end{figure}
\vfill\clearpage

\begin{figure}
\figurenum{8a}
\plotone{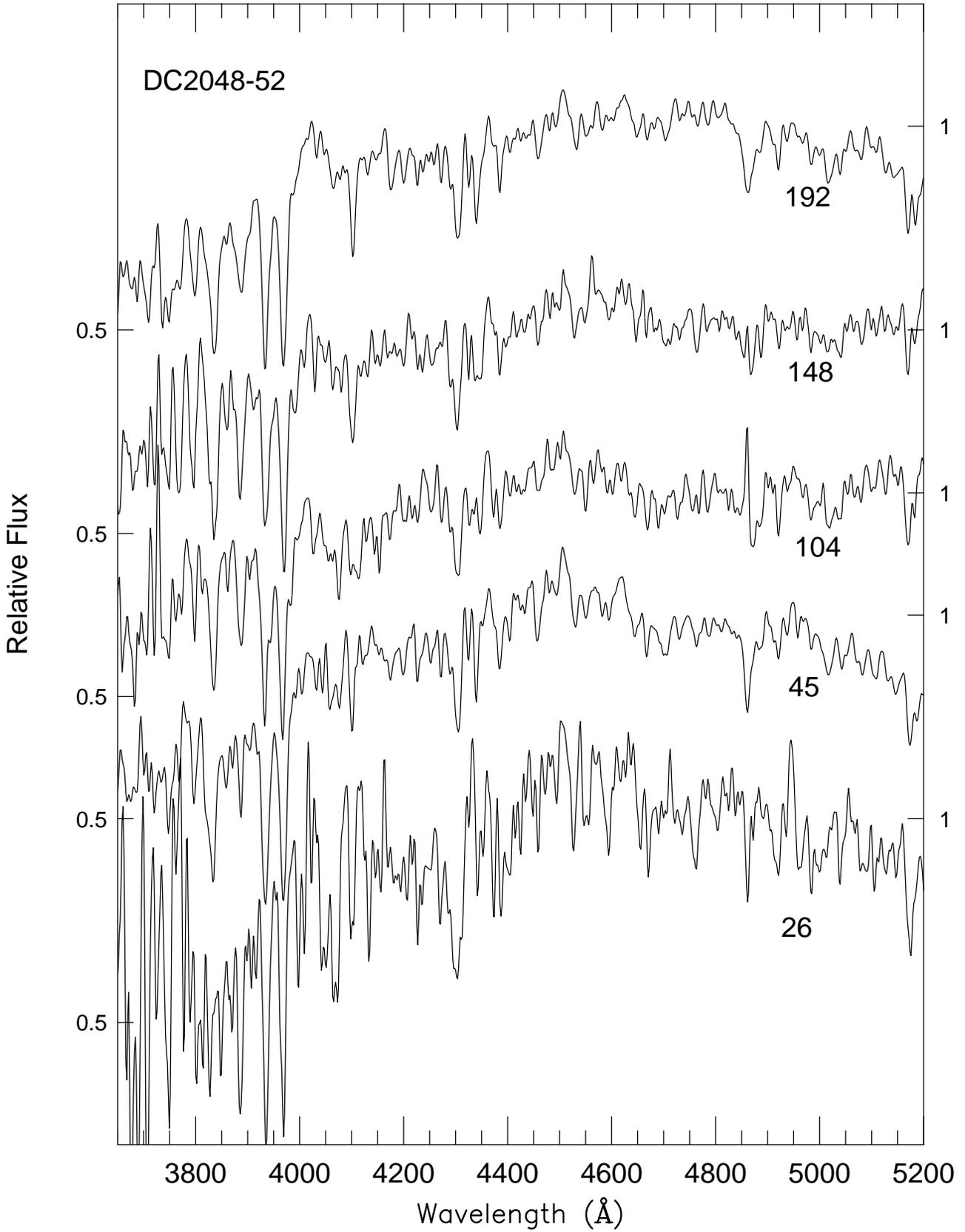}
\caption{ Spectra in DC2048--52: PSB galaxies without emission.}
\end{figure}
\vfill\clearpage

\begin{figure}
\figurenum{8b}
\plotone{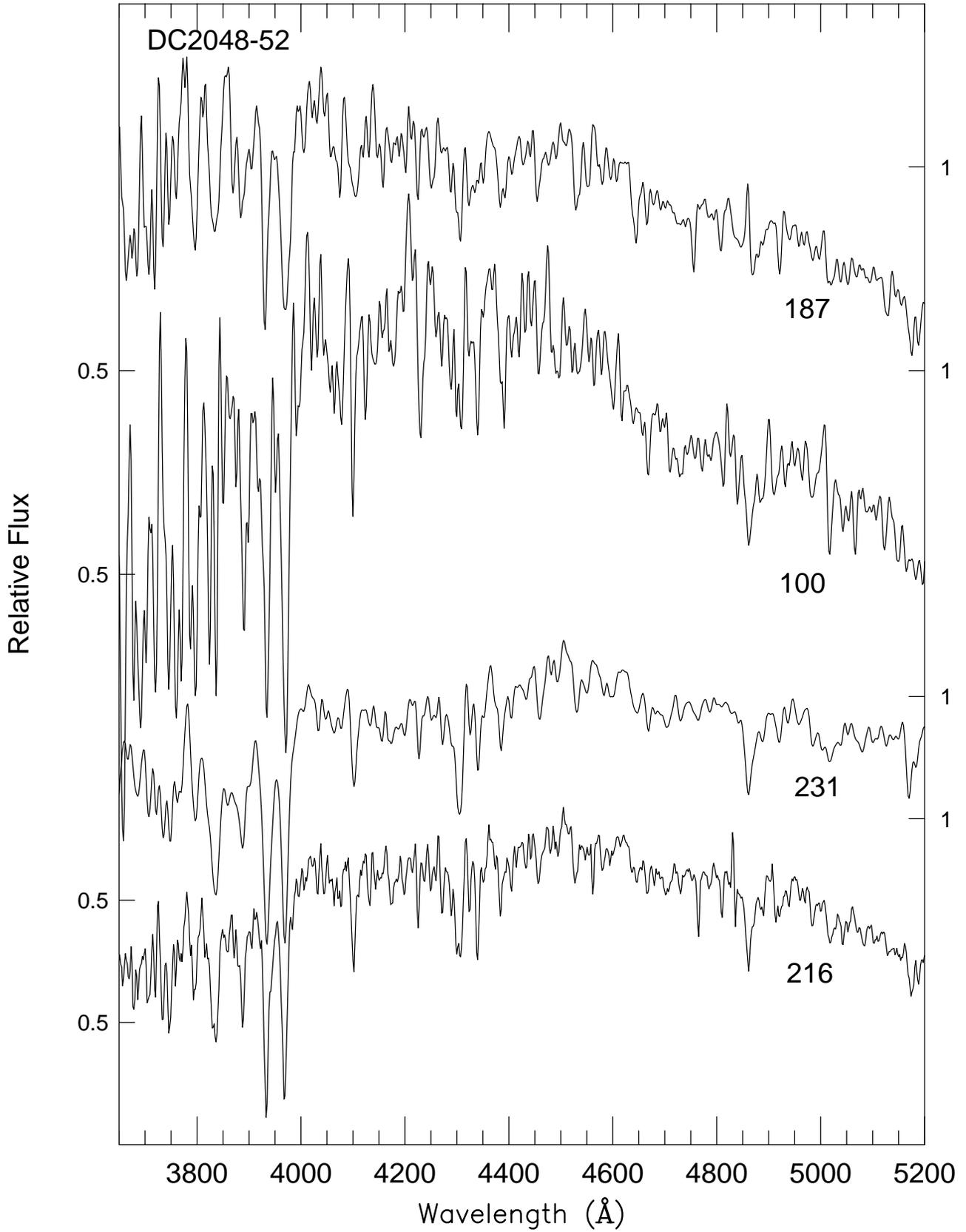}
\caption{DC2048--52 PSB galaxies with some emission.}
\end{figure}
\vfill\clearpage

\begin{figure}
\figurenum{8c}
\plotone{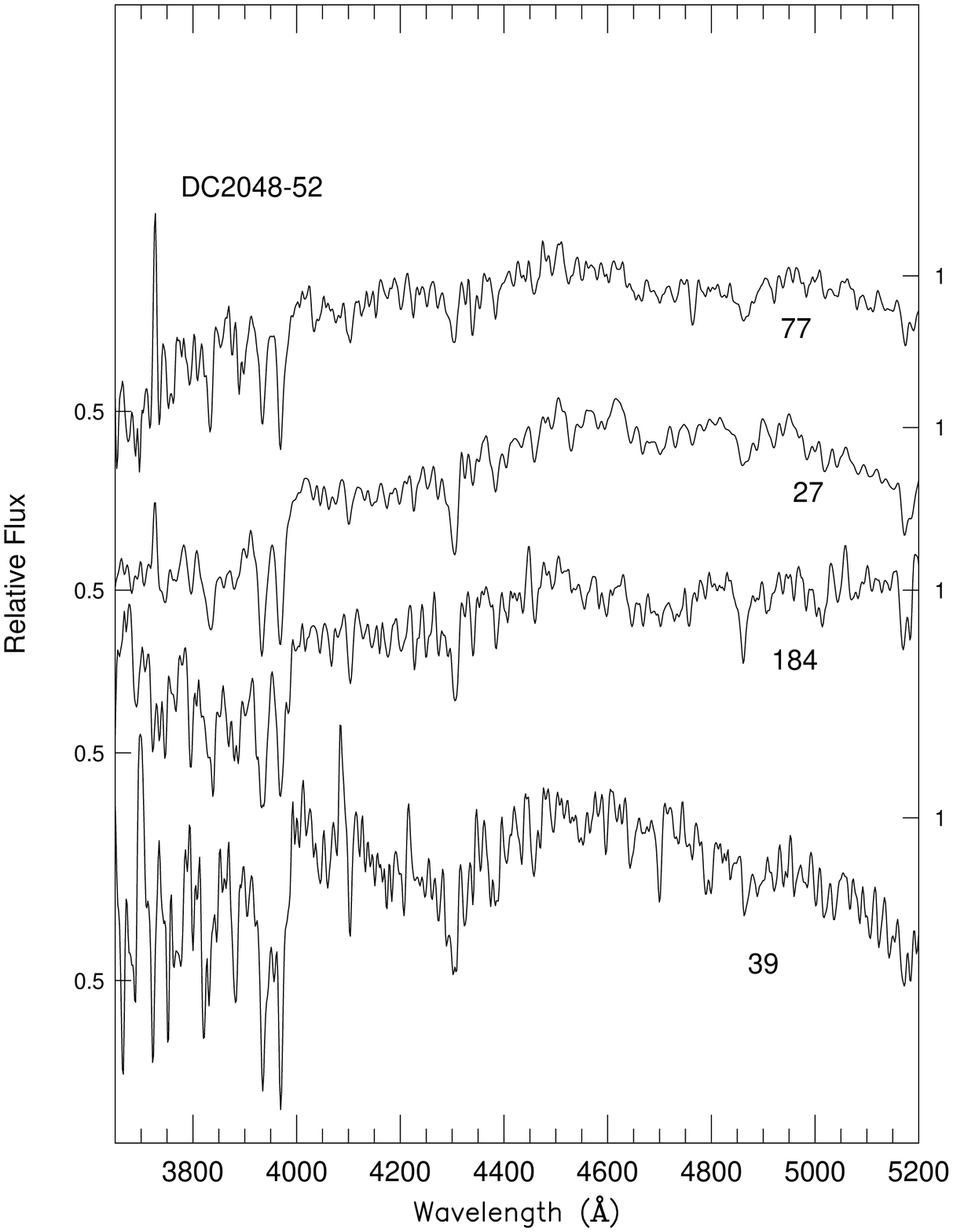}
\caption{same as (b).}
\end{figure}
\vfill\clearpage

\begin{figure}
\figurenum{8d}
\plotone{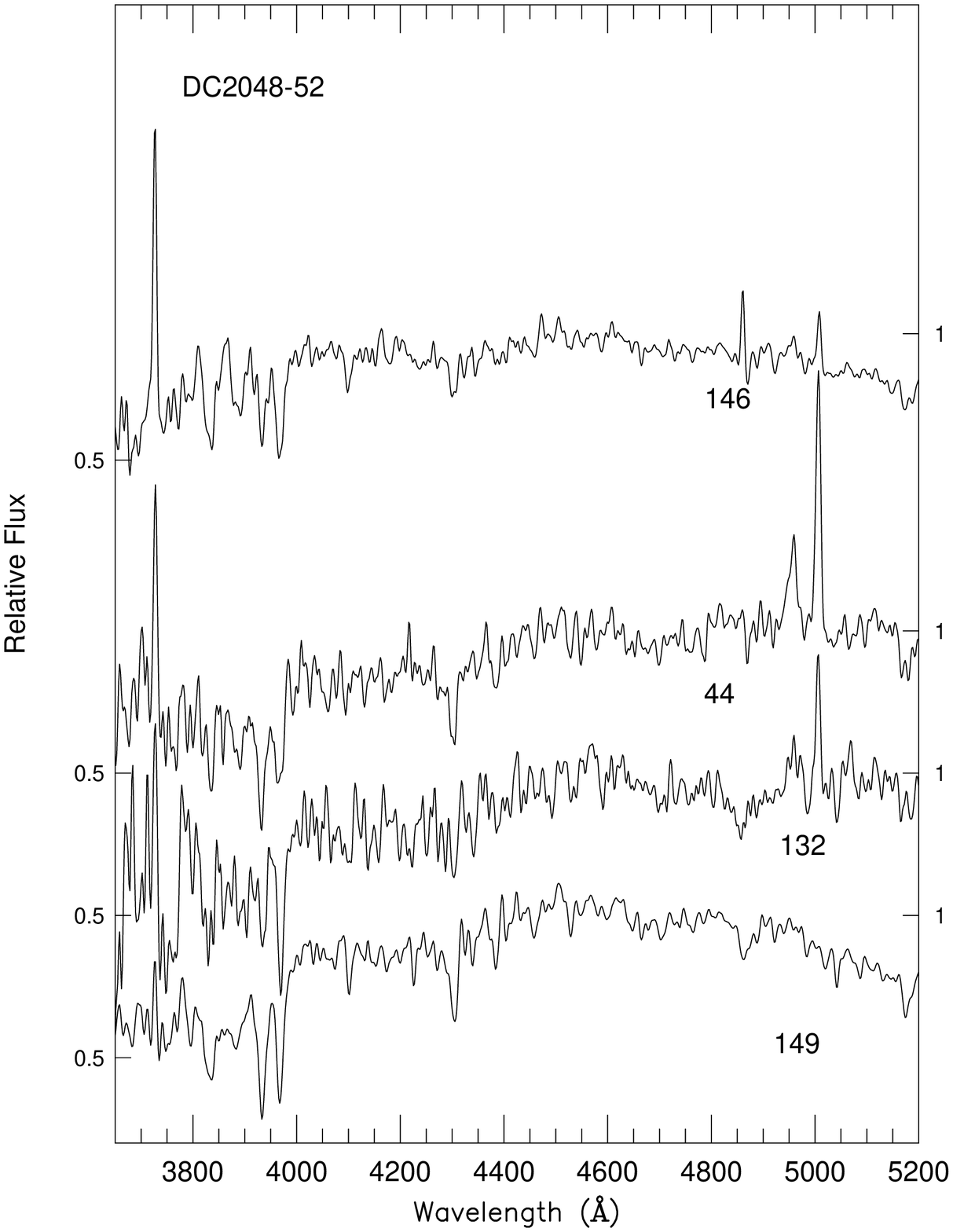}
\caption{strong emission line galaxies.}
\end{figure}
\vfill\clearpage

\begin{figure}
\figurenum{8e}
\plotone{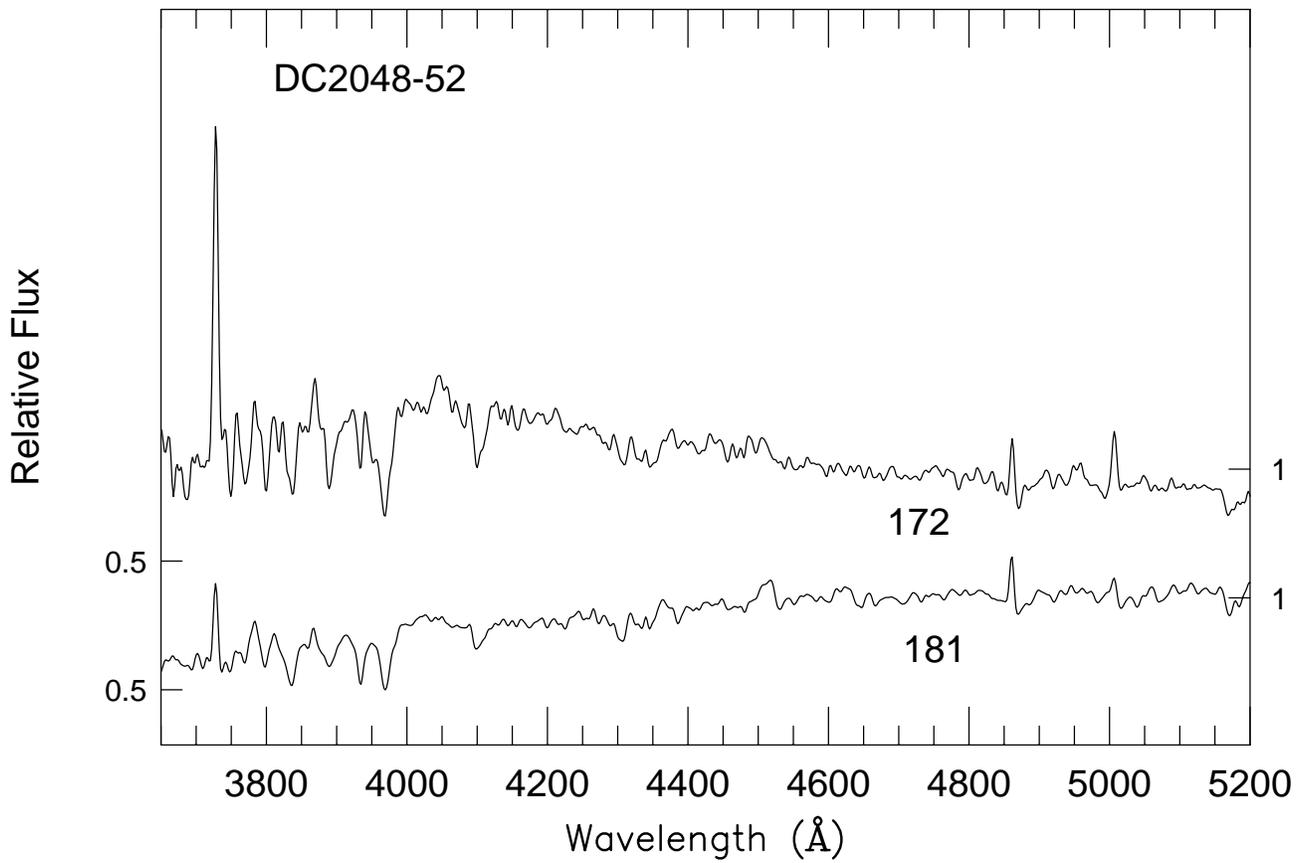}
\caption{same as (d).}
\end{figure}
\vfill\clearpage

\begin{figure}
\figurenum{9a}
\plotone{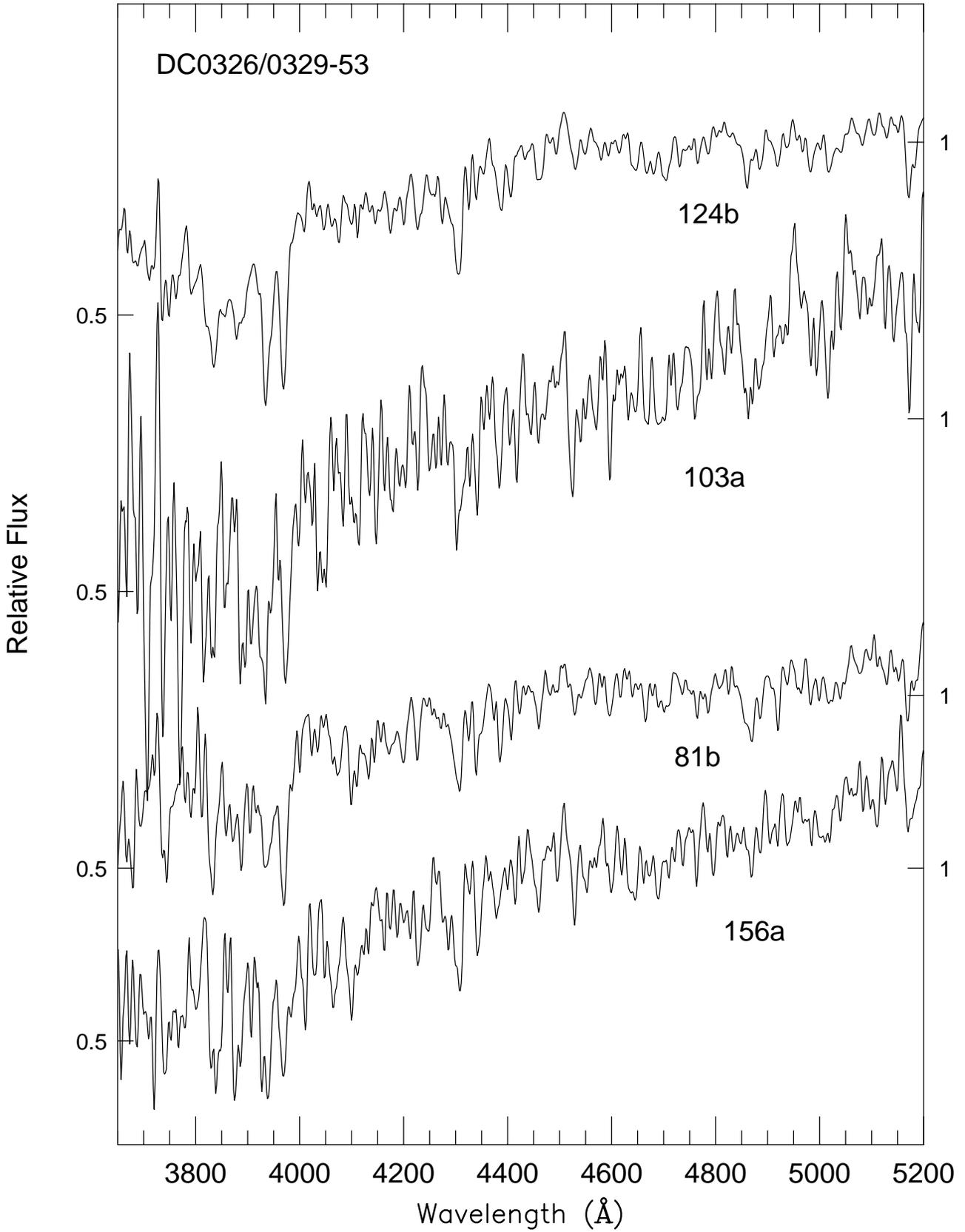}
\caption{PSB Spectra in DC0326--53/0329--52. Spectra with designations
``a'' are from DC0329--52, those with ``b'' are from DC0326--53.}
\end{figure}
\vfill\clearpage

\begin{figure}
\figurenum{9b}
\plotone{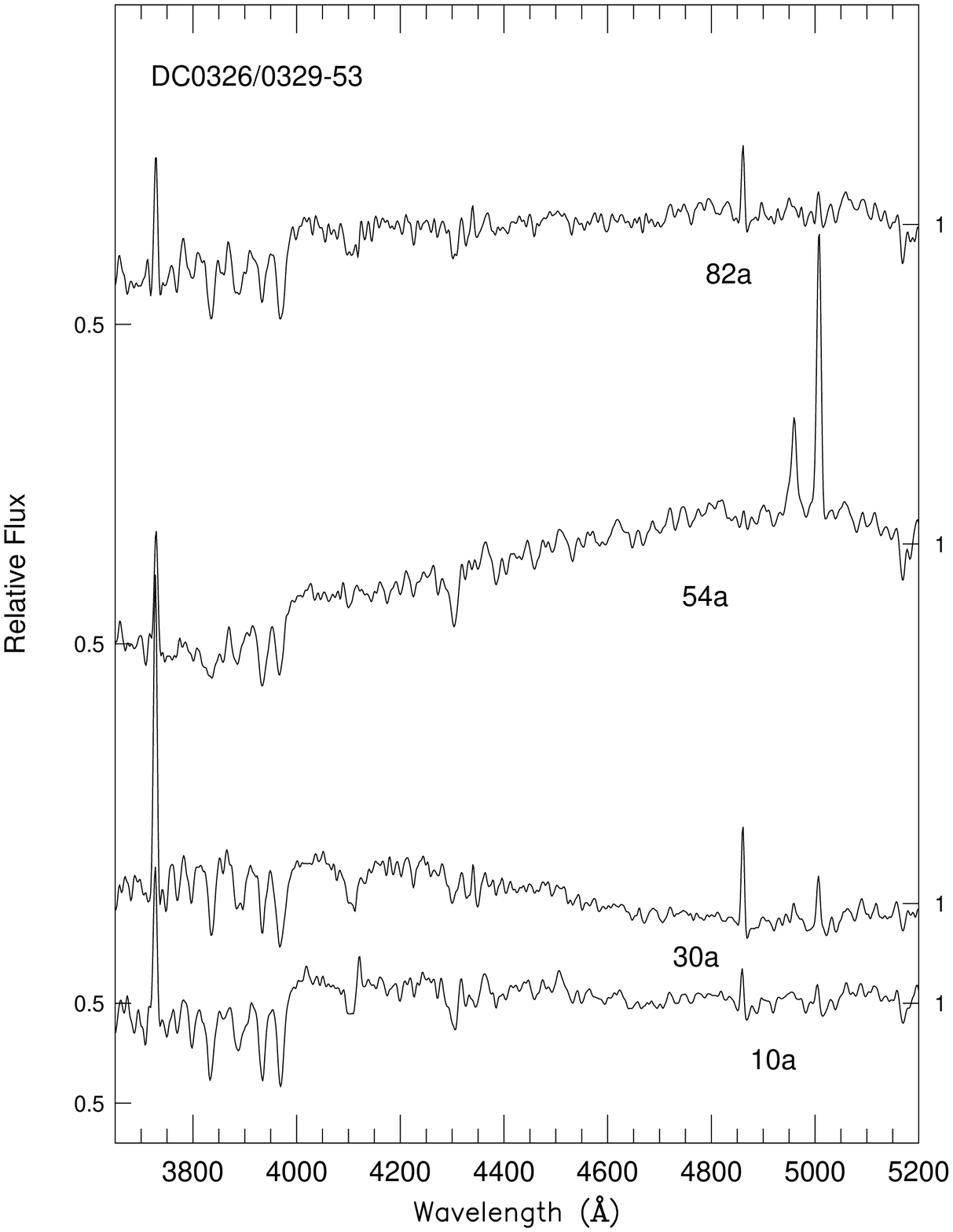}
\caption{strong emission line galaxies.}
\end{figure}
\vfill\clearpage

\begin{figure}
\figurenum{9c}
\plotone{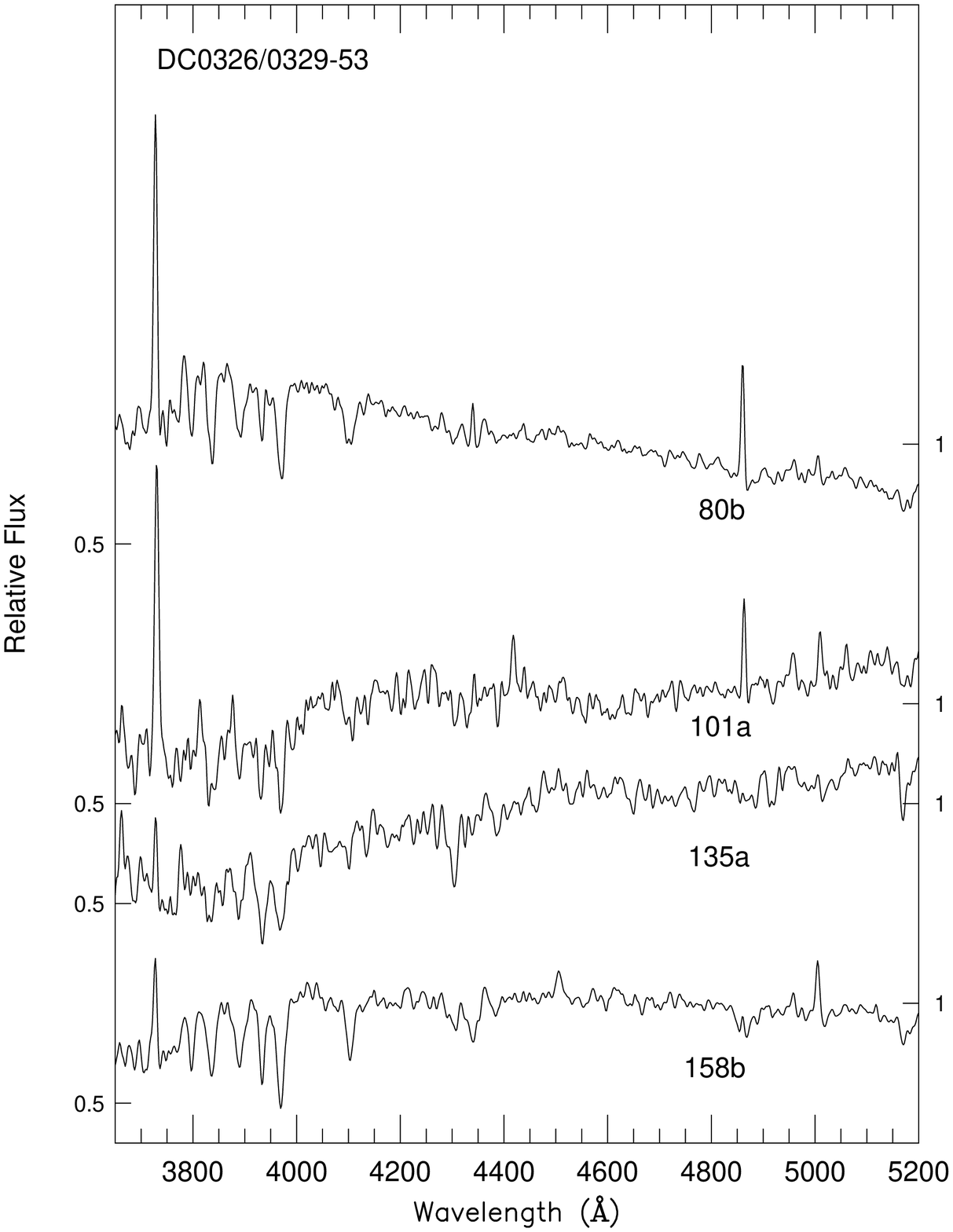}
\caption{ same as b.}
\end{figure}
\vfill\clearpage

\begin{figure}
\figurenum{10a}
\plotone{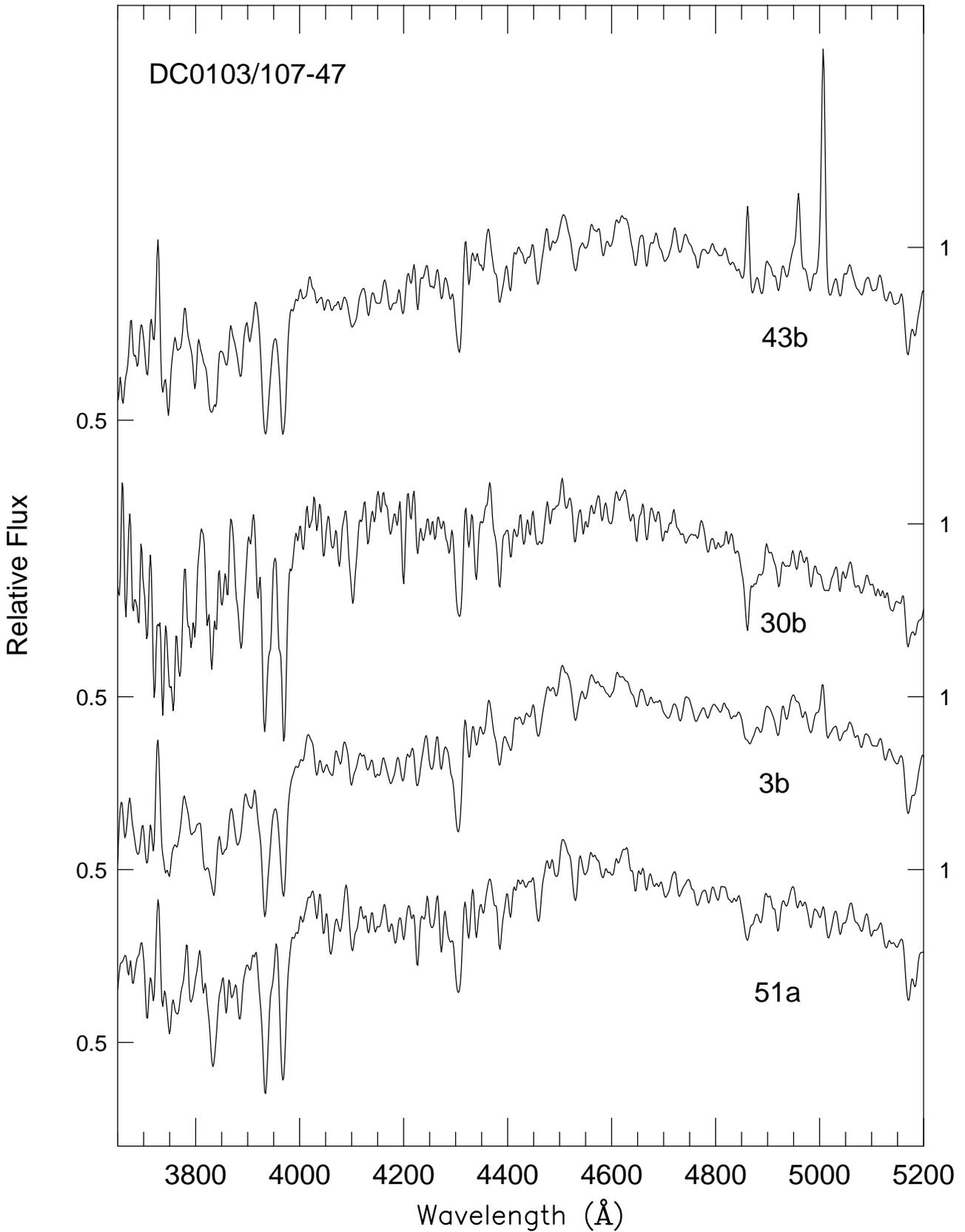}
\caption{PSB Spectra in DC0103--47/0107--46. Spectra with designations ``a'' are
from DC0103--47, those with ``b'' are from DC0107--46, those with ``m'' from 
table 4 in Malumuth et al. (1992).}
\end{figure}
\vfill\clearpage

\begin{figure}
\figurenum{10b}
\plotone{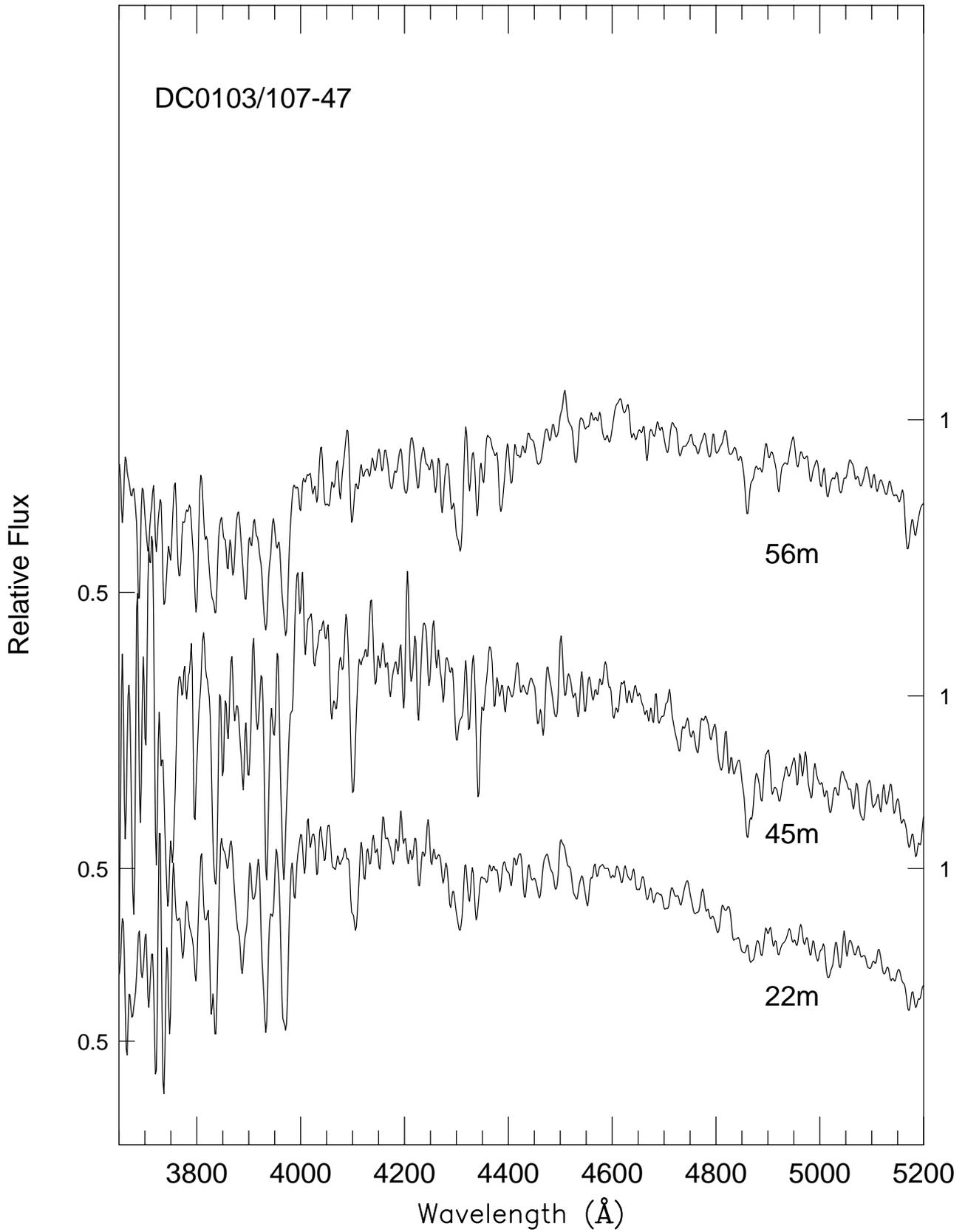}
\caption{same as (a).}
\end{figure}
\vfill\clearpage

\begin{figure}
\figurenum{11}
\plotone{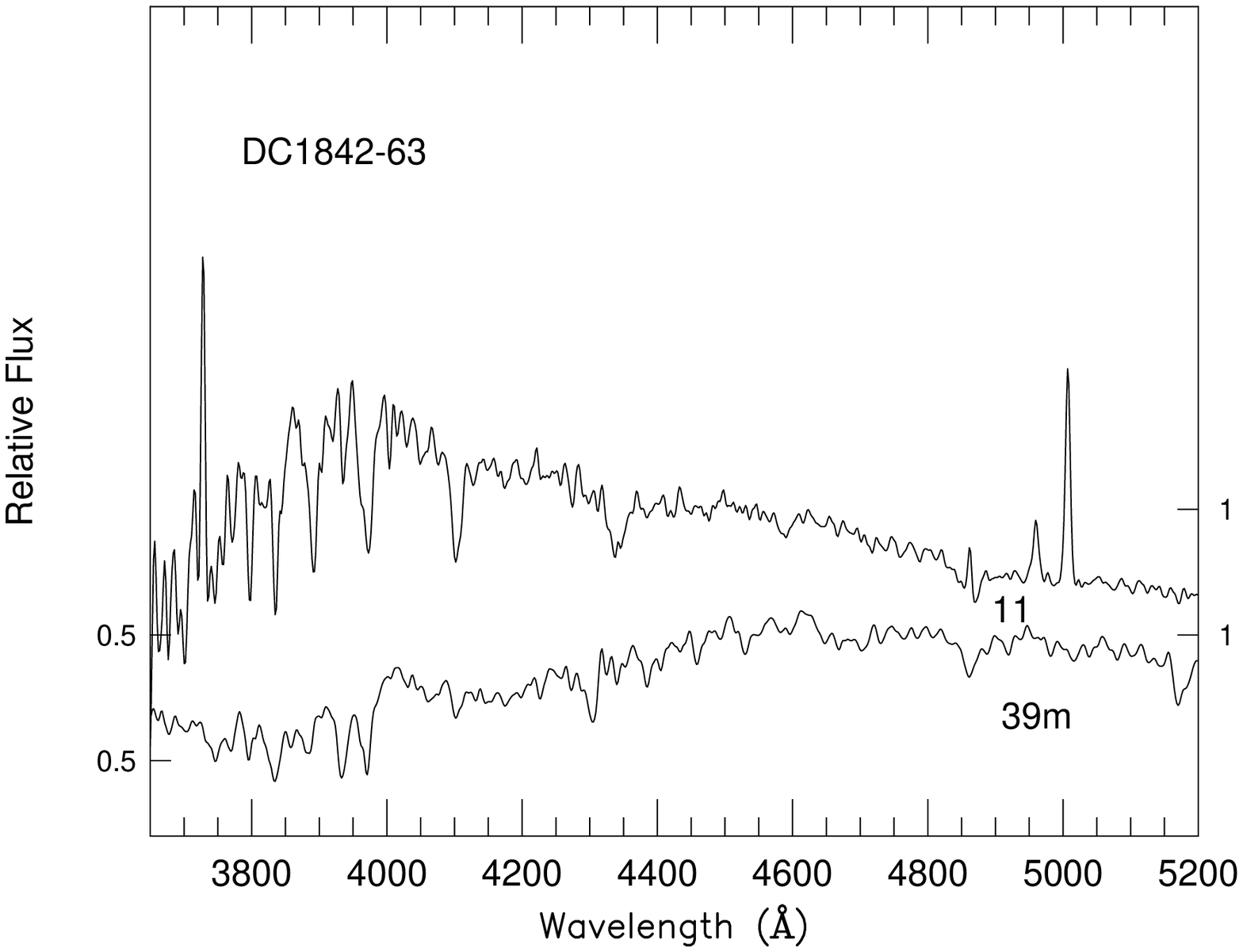}
\caption{Spectra in DC1842--63. ``m'' refers to table 7 of Malumuth et al. (1992).}
\end{figure}
\vfill\clearpage

%\begin{figure}
%\figurenum{12a}
%\plotone{caldwell.fig12a.ps}
%\caption{Images of many of the abnormal spectrum galaxies in the five clusters 
%studied.  Most of these blue CCD images were taken with the CTIO 0.9m;
%the Coma galaxies were observed with the FLWO 1.2m telescope.  Coma and 
%DC0103--46/0107--47, \#61 in Coma is shown in U and V.}
%\end{figure} 
%\vfill\clearpage

%\begin{figure}
%\figurenum{12b}
%\plotone{caldwell.fig12b.ps}
%\caption{DC2048--52.}
%\end{figure} 
%\vfill\clearpage

%\begin{figure}
%\figurenum{12c}
%\plotone{caldwell.fig12c.ps}
%\caption{DC0326--52/0329-53, terminology as in the tables.}
%\end{figure} 
%\vfill\clearpage

\begin{figure}
\figurenum{13}
\plotone{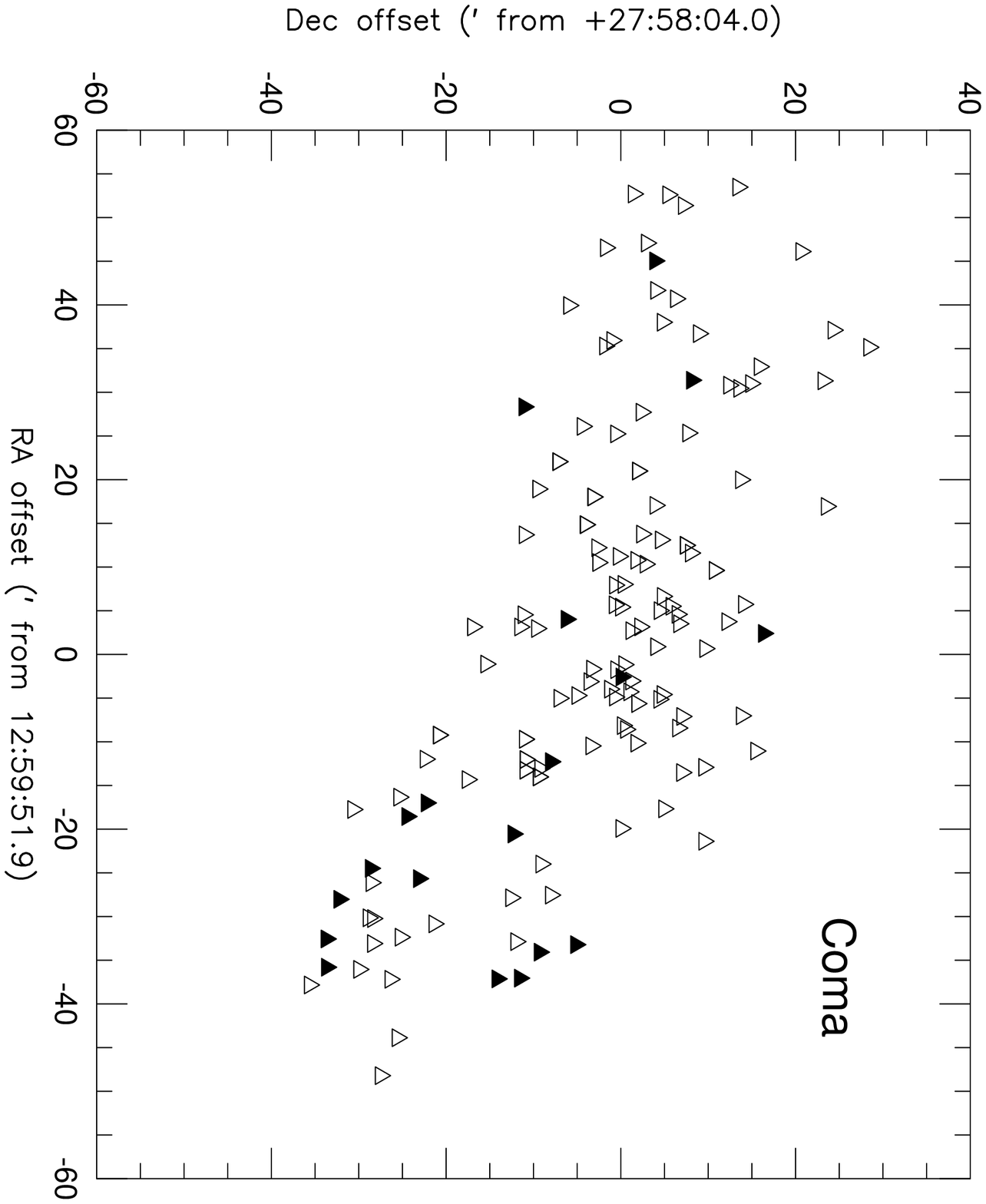}
\caption{Positions of normal (open triangles) and abnormal (filled triangles) 
spectrum galaxies in the Coma cluster with B$\leq$17.5.}
\end{figure}
\vfill\clearpage

\begin{figure}
\figurenum{14}
\plotone{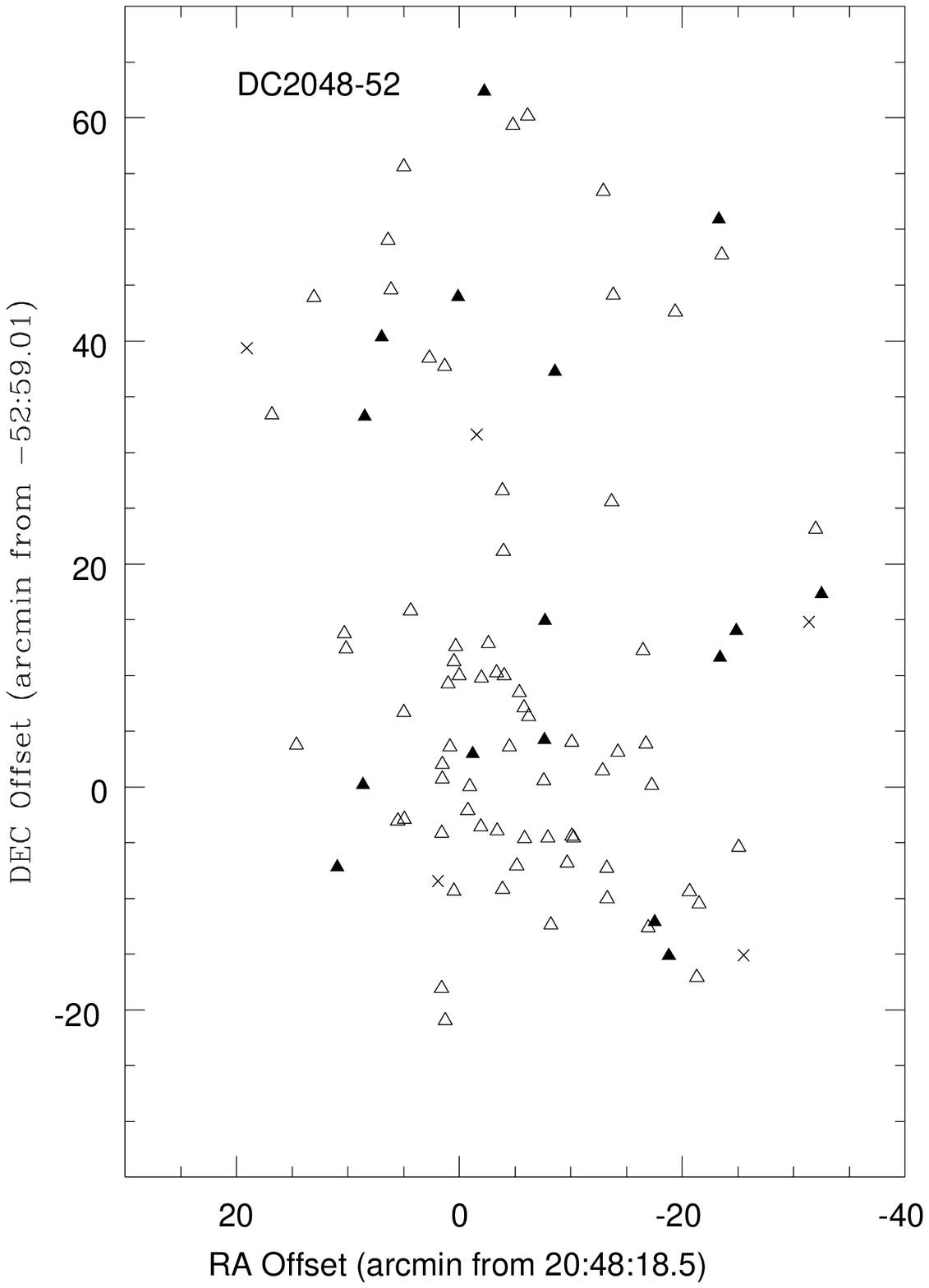}
\caption{Positions of normal and abnormal spectrum galaxies
in DC2048--52.  Same symbols as in Fig. 13.}
\end{figure}
\vfill\clearpage

\begin{figure}
\figurenum{15}
\plotone{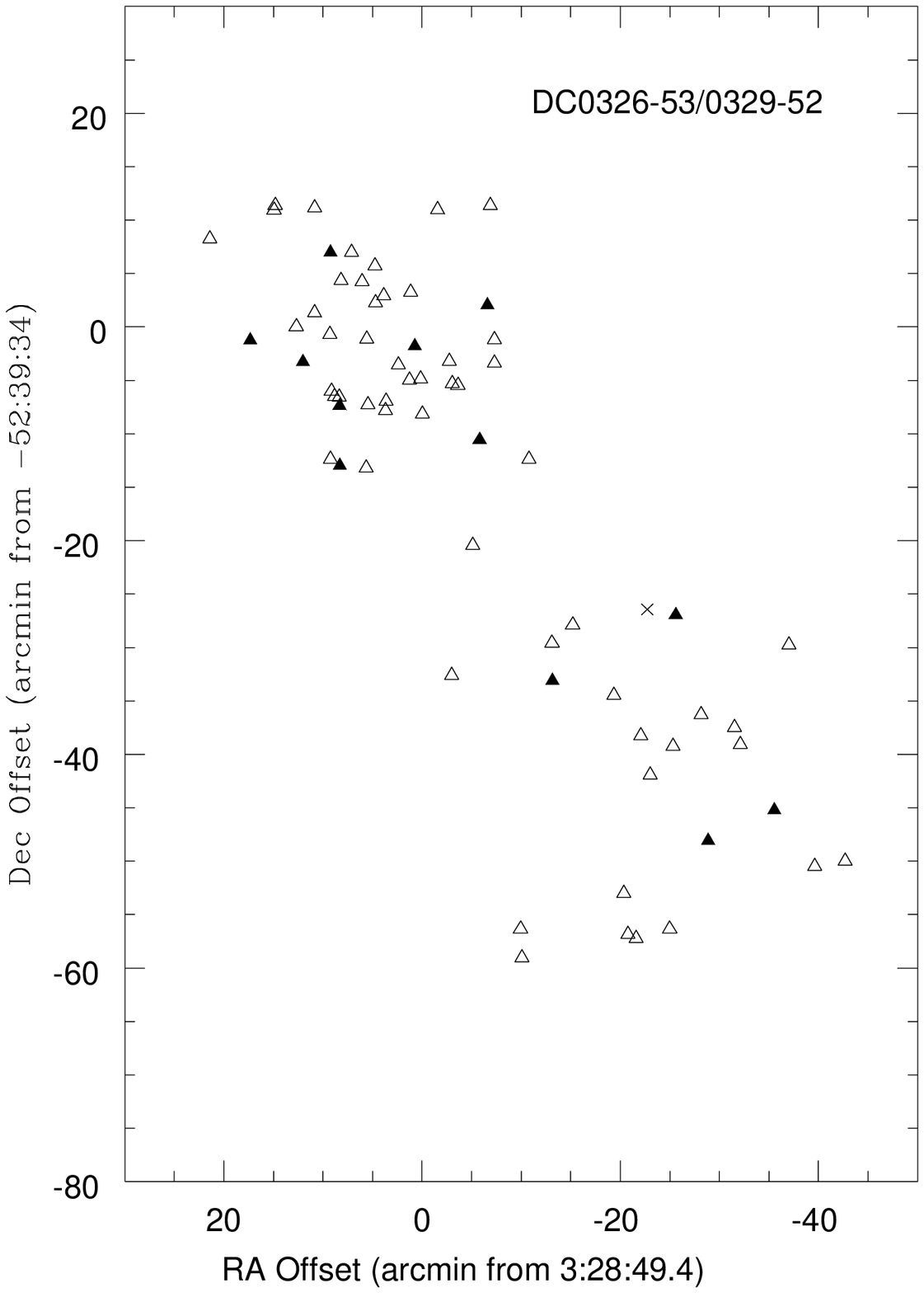}
\caption{Positions of normal and abnormal spectrum galaxies
in DC0326--53/0329--52.  Same symbols as in Fig. 13.}
\end{figure}
\vfill\clearpage

\begin{figure}
\figurenum{16}
\plotone{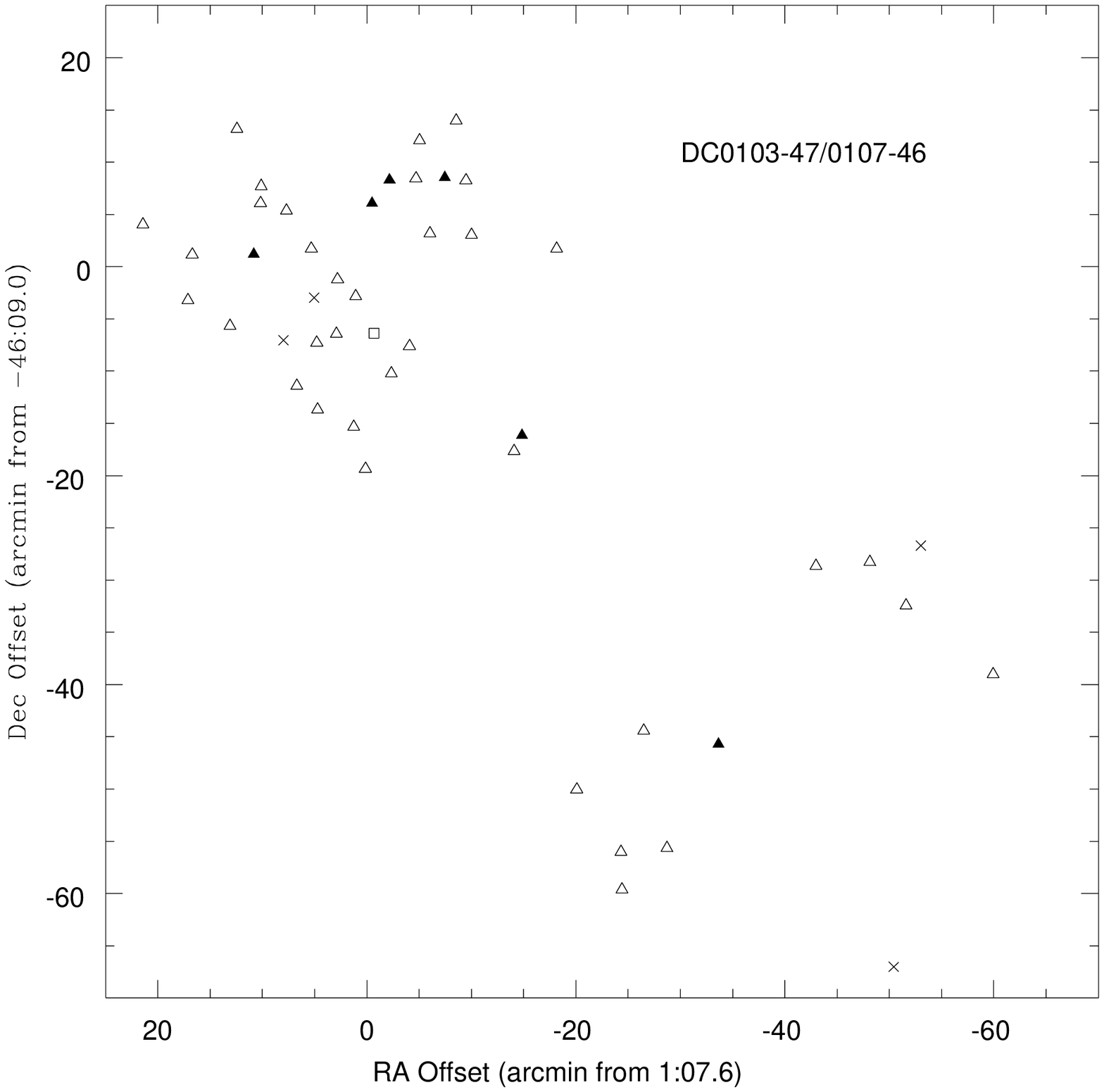}
\caption{Positions of normal and abnormal spectrum galaxies
in DC0103--47/0107--46.  Same symbols as in Fig. 13.}
\end{figure}
\vfill\clearpage

\begin{figure}
\figurenum{17}
\plotone{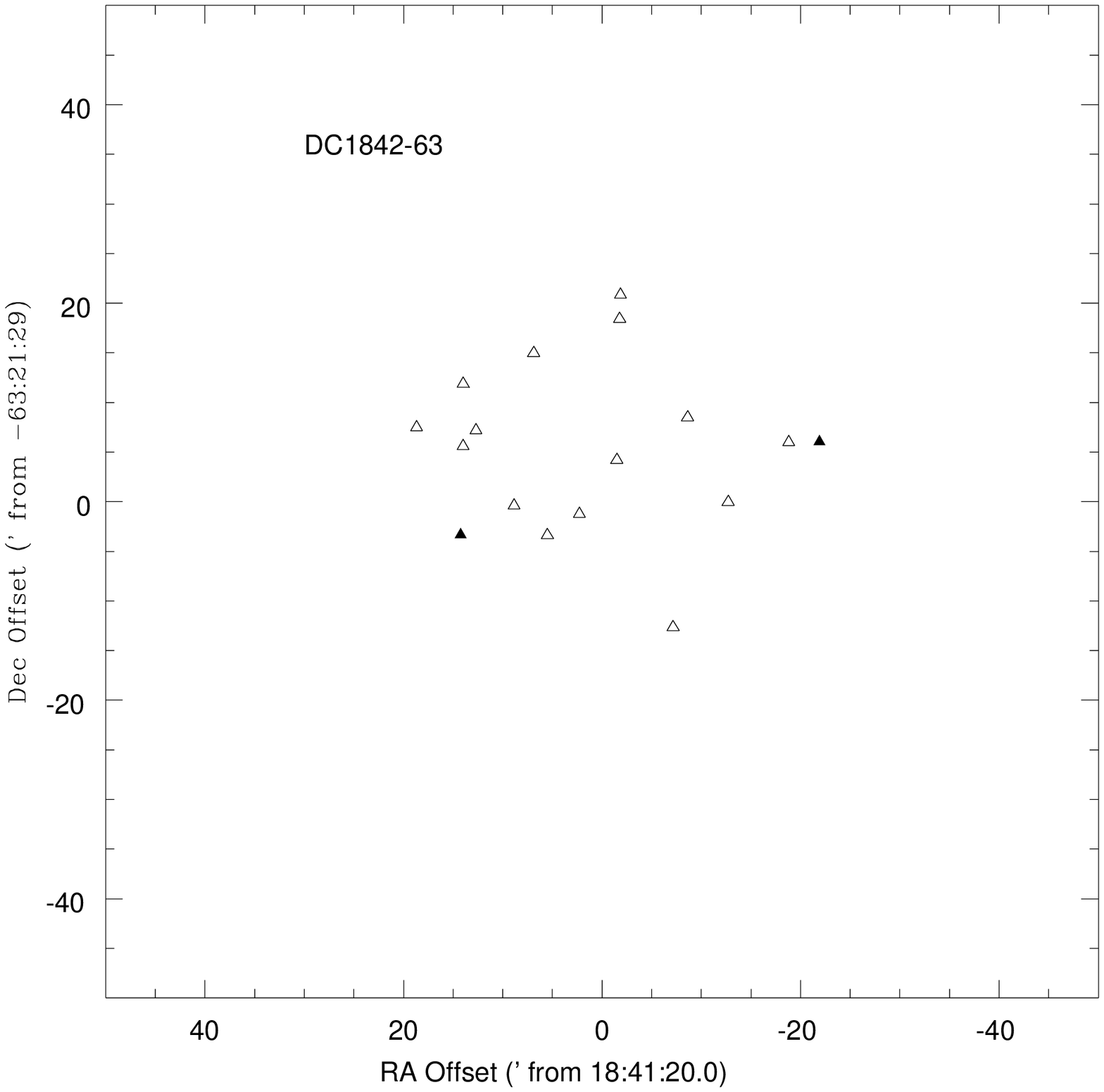}
\caption{Positions of normal and abnormal spectrum galaxies
in DC1842--63.  Same symbols as in Fig. 13.}
\end{figure}
\vfill\clearpage

\begin{figure}
\figurenum{18}
\plotone{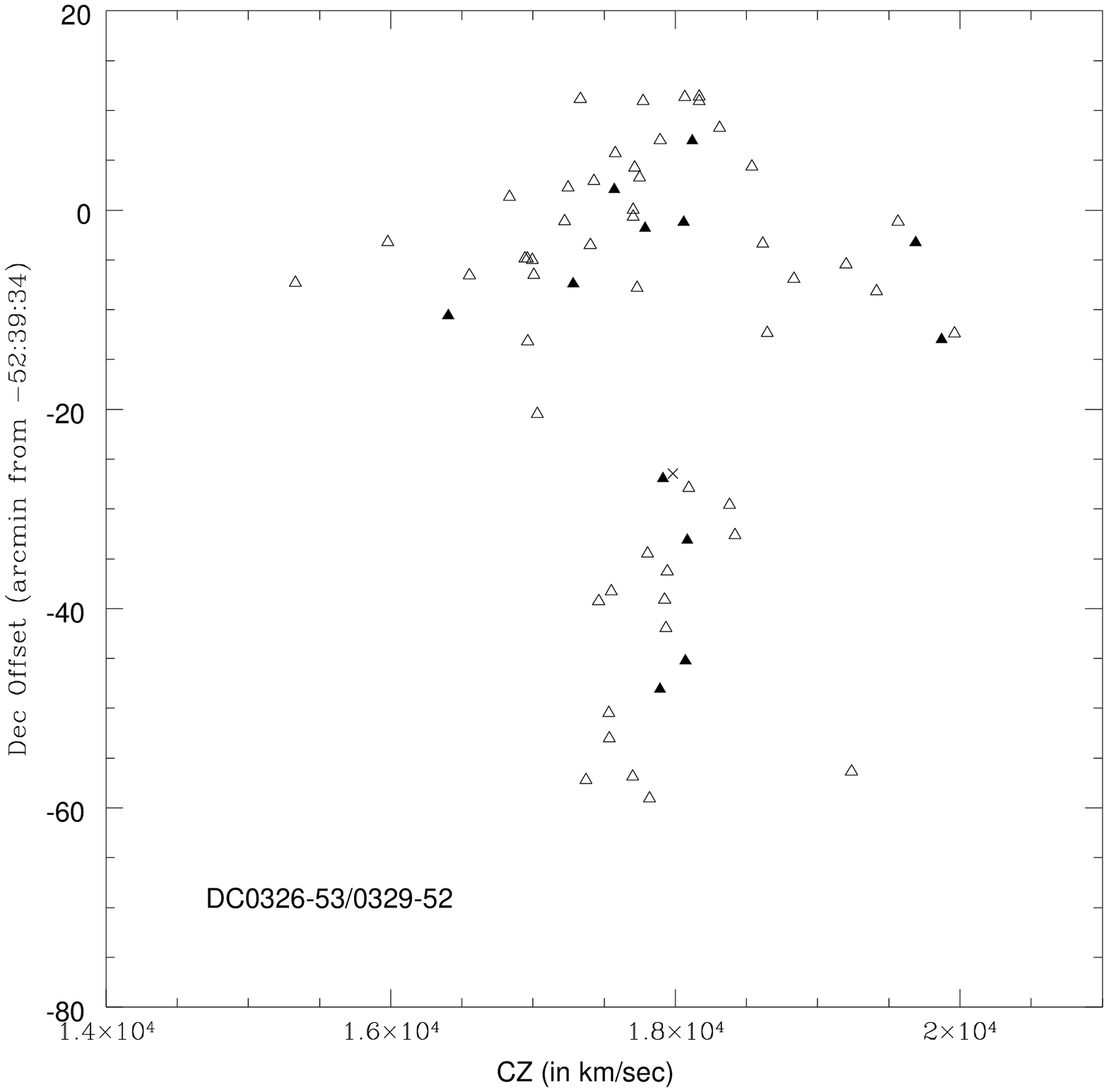}
\caption{Dec versus cz plot for normal (unfilled squares) and
abnormal (filled circles) spectrum galaxies in DC0326--53/0329--52.  Galaxies for
which the spectra are inconclusive are plotted as x's.}
\end{figure}
\vfill\clearpage

\begin{figure}
\figurenum{19}
\plotone{caldwell.fig19.psx}
\caption{Positions of DC0329--52 galaxies observed by us spectroscopically (x's)
are overlaid on x-ray emission contours from the ROSAT archival PSPC image.
The contours show the double peak x-ray structure commented on in the text.}
\end{figure}
\vfill\clearpage

\begin{figure}
\figurenum{20}
\plotone{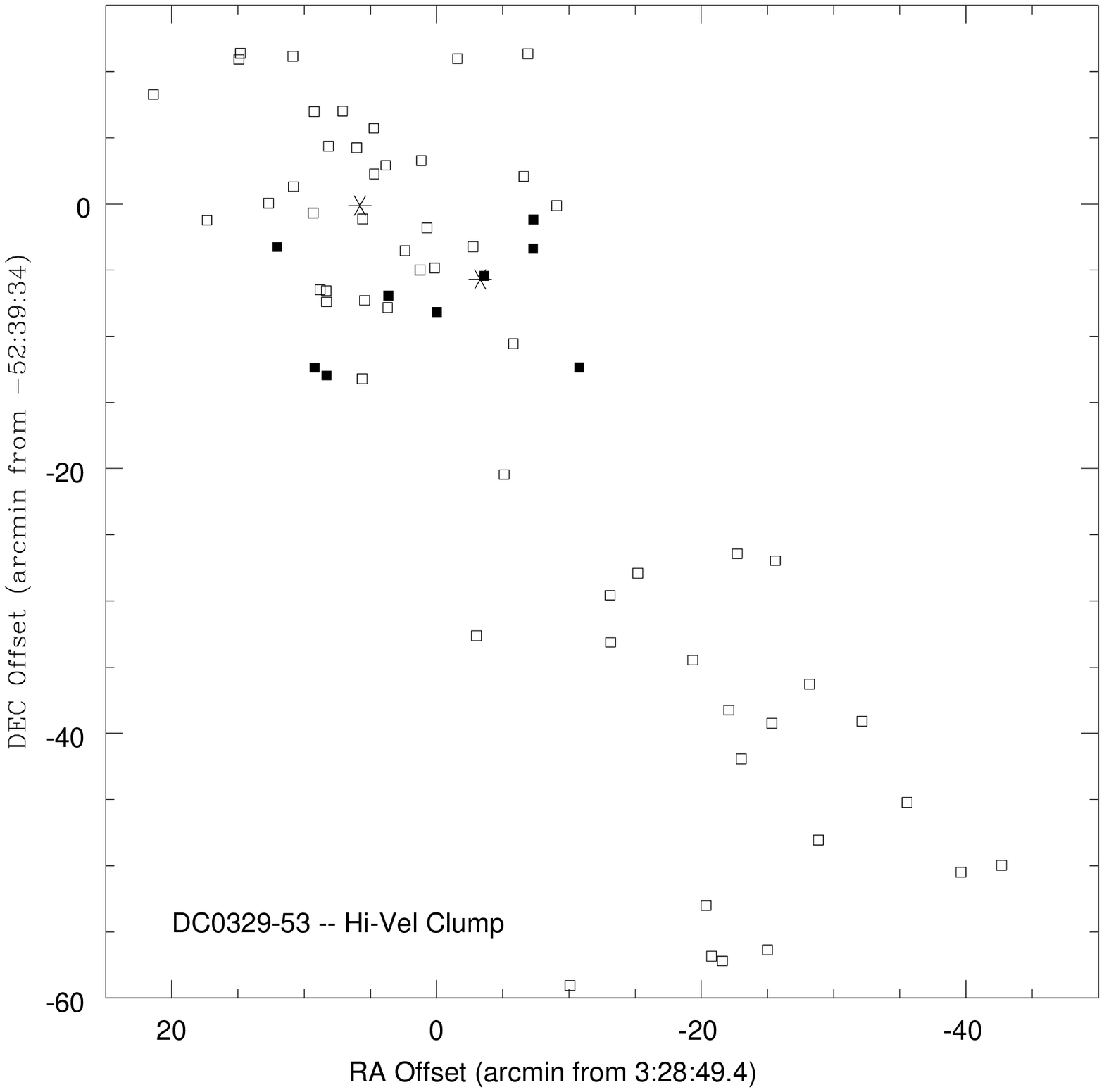}
\caption{Positions of the high-velocity clump galaxies (filled squares)
in DC0329--52 are compared to those of the other galaxies in DC0329--52 (unfilled
squares) and to the location of the two ROSAT x-ray emission peaks (large
plusses).  The brighter x-ray peak is the one to the SW.}
\end{figure}
\vfill\clearpage

\begin{figure}
\figurenum{21}
\plotone{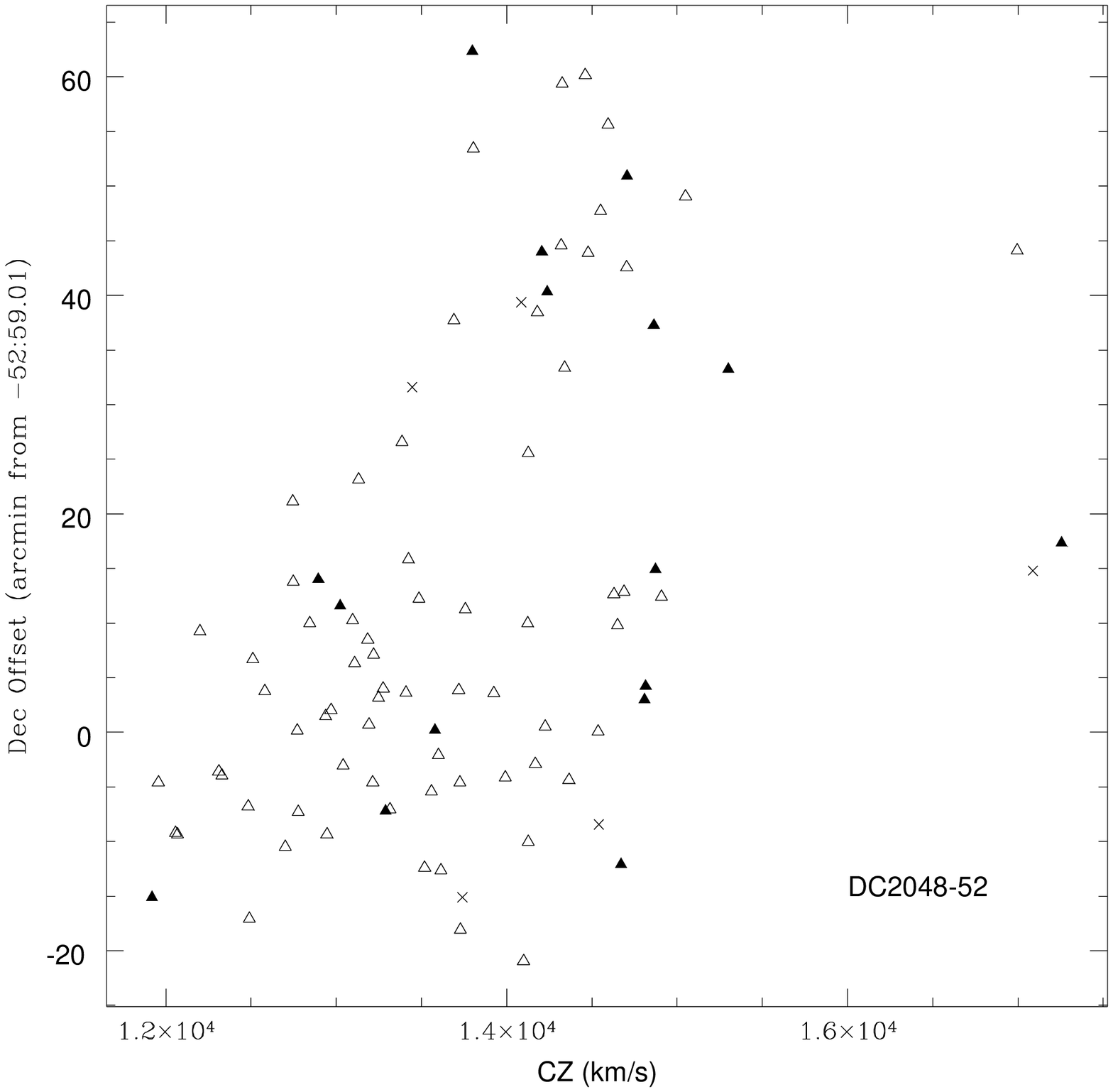}
\caption{Dec versus redshift (cz) plot for galaxies in DC2048--52.  Same symbols
as in Fig. 18.}
\end{figure}
\vfill\clearpage

\begin{figure}
\figurenum{22}
\plotone{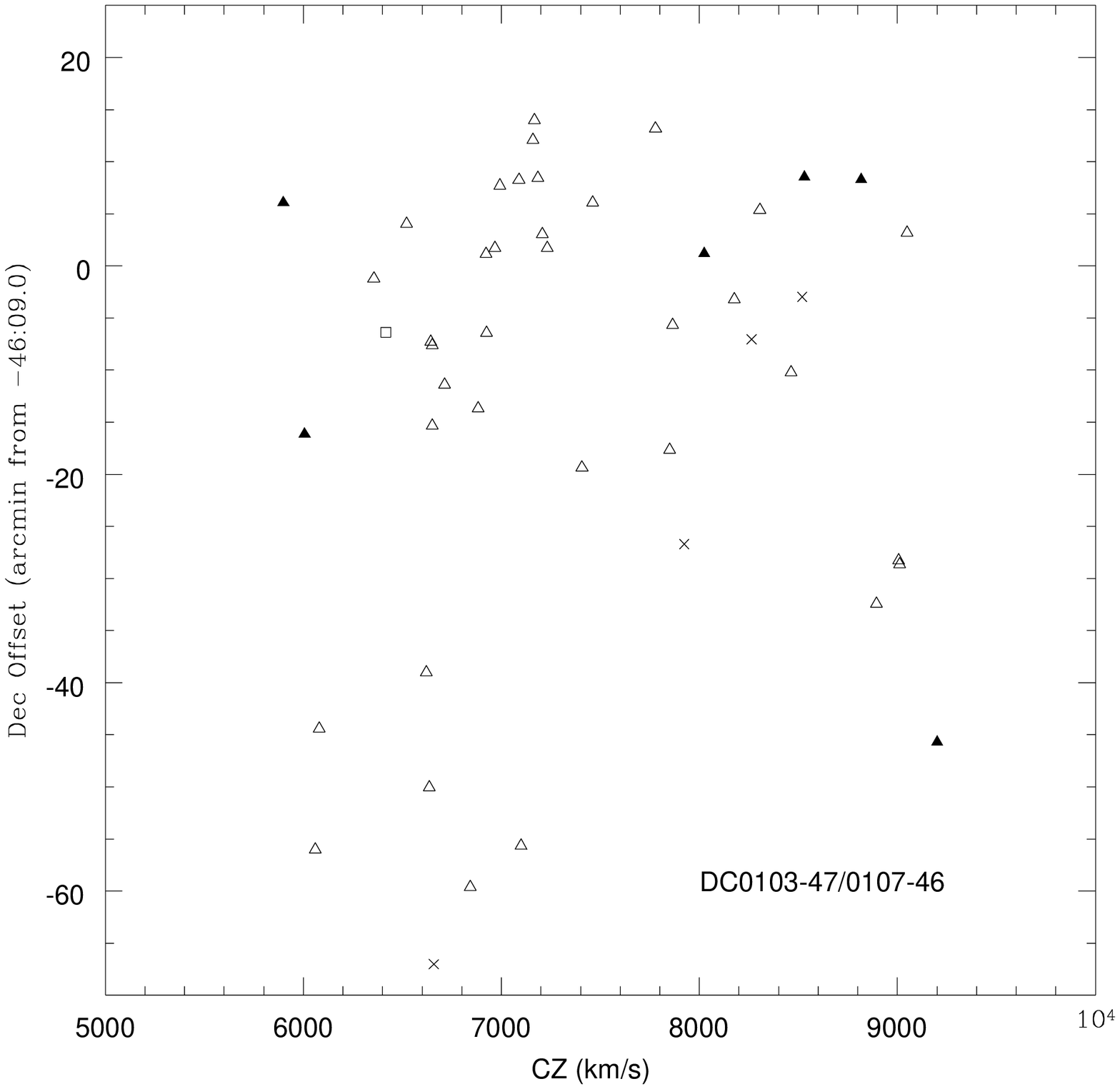}
\caption{Dec versus redshift (cz) plot for galaxies in DC0103-47/0107--46.  Same
symbols as in Fig. 18.}
\end{figure}
\vfill\clearpage

\begin{figure}
\figurenum{23}
\plotone{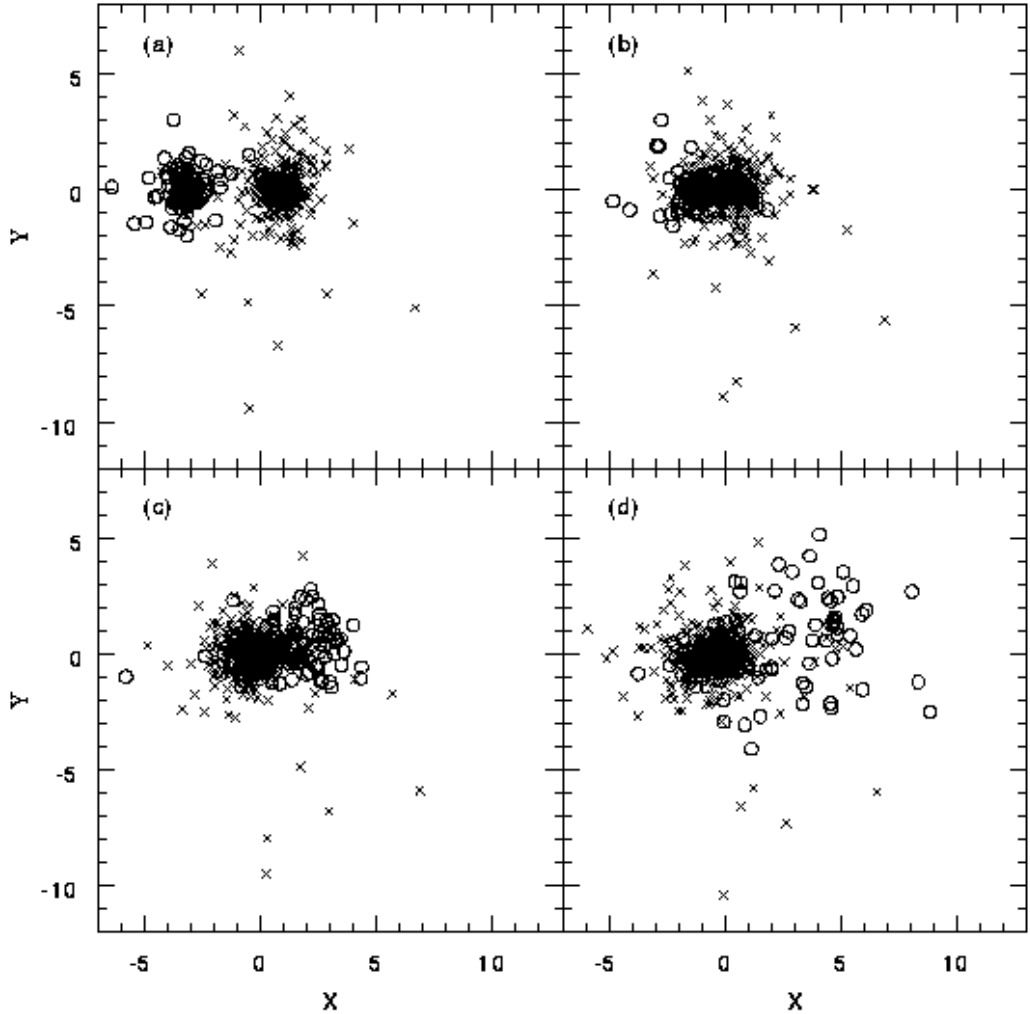}
\caption{N-body simulation of a subcluster of galaxies (open circles) passing 
through the main cluster (x's); the cluster to subcluster mass ratio is 4:1.
The X and Y positions are in units of the mean
harmonic radius of the main cluster.  The simulation begins at (a) with the
subcluster at a distance of 4 mean harmonic radii along the X axis, and with 
an infall velocity along the X axis that is one fourth of parabolic.  The time 
steps in (b), (c), and (d)
are 5, 10, and 15 time units later, respectively, where each time unit is about
0.4 dynamical timescales.}
\end{figure}
\vfill\clearpage

\begin{figure}
\figurenum{24}
\plotone{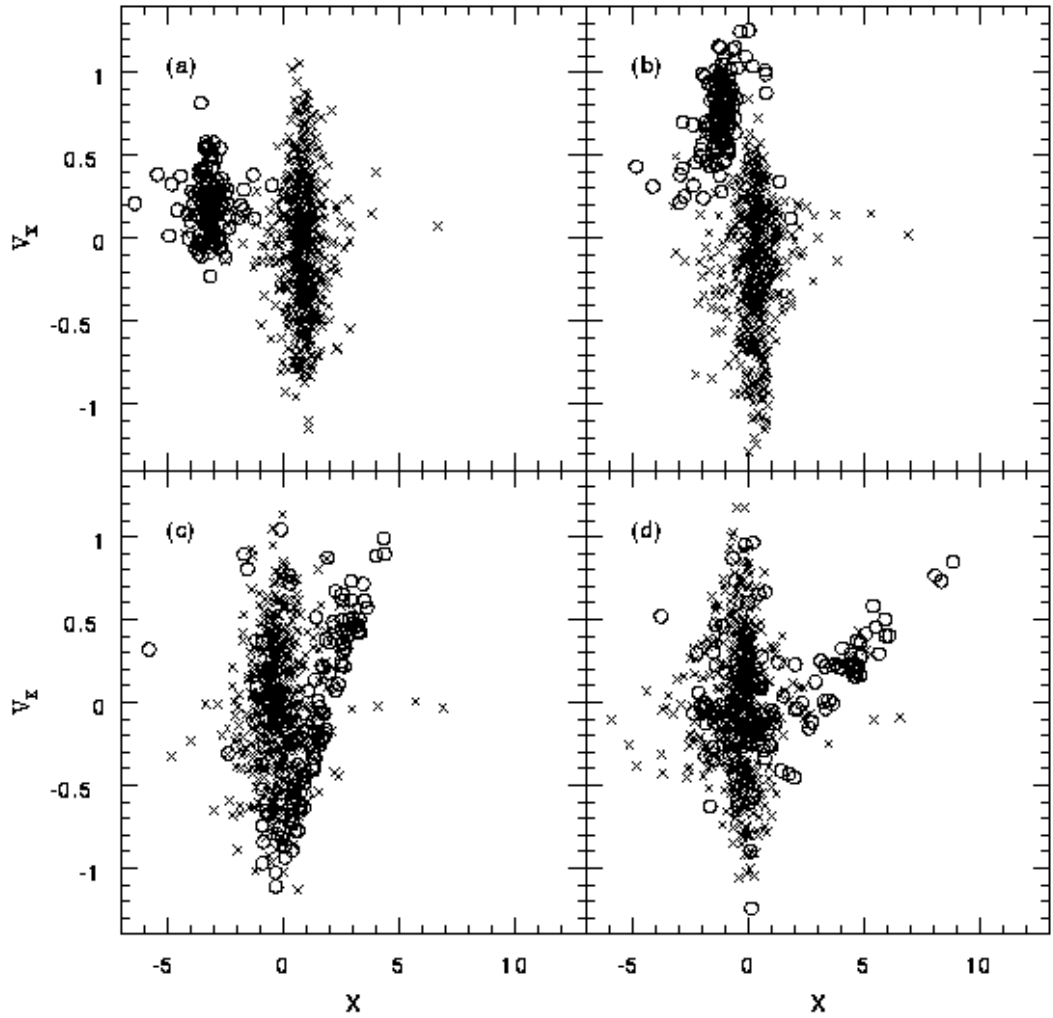}
\caption{X versus V$_x$ plot for the same simulation as in Fig. 20 and at the
same four time steps.  As can be seen in the Figure, a strong tidal distension
of the subcluster occurs only after the subcluster passes through the main cluster.}
\end{figure}
\vfill\clearpage

% That's all, folks.
%
% The technique of segregating major semantic components of the document
% within "environments" is a very good one, but you as an author have to
% come up with a way of making sure each \begin{whatzit} has a corresponding
% \end{whatzit}.  If you miss one, LaTeX will probably complain a great
% deal during the composition of the document.  Occasionally, you get away
% with it right up to the \end{document}, in which case, you will see
% "\begin{whatzit} ended by \end{document}".

\end{document}